\documentclass[12pt]{article}
\usepackage{graphicx}
\usepackage{geometry}
\usepackage{mathtools}
\usepackage{xcolor}
\usepackage{cancel}
\usepackage{amssymb}
\usepackage{mathrsfs}
\usepackage{soul}

\newcommand{\snot[1]}{{\scriptstyle \times 10}^{#1}}

 \def\norm#1{\vert #1 \vert}
 \def\Norm#1{\Vert #1\Vert}

\title{Transits close to the Lagrangian solutions $L_1,L_2$ 
in the Elliptic Restricted Three-body Problem}
\author{
        Roc\'io I. Paez \& Massimiliano Guzzo\\
        {\small Dipartimento di Matematica ``Tullio Levi-Civita''}\\
        {\small Universit\`a degli Studi di Padova}\\
        {\small Padova 35122 (PD) Italia}}
\date{\today}

\begin{document}
\maketitle

\begin{abstract}
In the last decades a peculiar family of solutions of the Circular
Restricted Three Body Problem has been used to explain the temporary
captures of small bodies and spacecrafts by a planet of the Solar
System. These solutions, which transit close to the Lagrangian points
$L_1,L_2$ of the CRTBP, have been classified using the values of approximate
local integrals and of the Jacobi constant. The use for small bodies of the
Solar System requires to consider a hierarchical extension of the
model, from the CRTBP to the the full $N$ planetary problem.  The
Elliptic Restricted Three Body, which is the first natural extension
of the CRTBP, represents already a challenge, since global first
integrals such as the Jacobi constant are not known for this
problem. In this paper we extend the classification of the transits
occurring close to the Lagrangian points $L_1,L_2$ of the ERTBP using
a combination of the Floquet theory and Birkhoff normalizations.
Provided that certain non-resonance conditions are satisfied, we
conjugate the Hamiltonian of the problem to an integrable normal form
Hamiltonian with remainder, which is used to define approximate local
first integrals and to classify the transits of orbits through a
neighbourhood of the Lagrange equilibria according to the values of
these integrals. We provide numerical demonstrations for the
Earth-Moon ERTBP.
\end{abstract} 

\section{Introduction}

The mathematics of the close encounters of a small body with a planet
has a long story, started with the discoveries of Lexell and Leverrier
that planet Jupiter can expel a comet from the Solar System
\cite{Leverrier}. Today a rich recent literature is concerned with
close encounters. See, for example,~\cite{FNS,GKZ,CG19a,CG19b} for
analytic studies, \cite{Lega11,GL13,GL16,GL17,GL18,PG20} for
applications to the dynamics of
comets,~\cite{Valsecchi,Valsecchi2,GT13,ABB17} for applications to
near-Earth asteroids dynamics
and~\cite{Conley67,Simo99,JM99,KLMR06,GomKoLoMaMaRoss,celletti1,ZCMD,OTY19}
for space mission design. The close encounters occurring with a
low relative velocity of the small body with respect to the Planet
are studied, since the paper~\cite{Conley67}, using a peculiar family
of solutions of the Circular Restricted Three Body Problem. These
solutions transit close to the Lagrangian points $L_1,L_2$ of the
CRTBP. But, the use for small bodies of the Solar System requires to
consider hierarchical extensions of the model, from the CRTBP to the
the full $N$ planetary problem, which are still subject of study
\cite{BGMO16,JB20,JJCR20,SG20}.

The Elliptic Restricted Three Body problem, which represents the first
natural extension of the CRTBP, introduces a big change in the
dynamics, since no matter how small is the value of the eccentricity,
global first integrals such as the Jacobi constant are not known. This
problem is defined by the motion of a body $P$ of infinitesimally
small mass moving in the gravity field generated by two massive bodies
$P_1$ and $P_2$ (the primary and secondary body respectively) which
move around their common center of mass according to the well known
elliptic solutions of the two-body problem. It is usual to represent
the motion of $P$ using a rotating-pulsating reference frame $(x,y,z)$
whose origin is in the center of mass of $P_1$ and $P_2$, the $z$ axis
is orthogonal to the motion of $P_1,P_2$, and the $x,y$ axes are
rotating-pulsating so that the primary and secondary bodies remain at
fixed locations on the horizontal axis $x$. With standard units of
measure, the Hamiltonian representing the motions of $P$ in this
pulsating-rotating frame is:
\begin{equation}\label{eq:ori3bp}
  \begin{aligned}
    h(x,y,z,p_x,p_y,p_z,& f;e)  =  \frac{p_x^2}{2} + \frac{p_y^2}{2}
    + \frac{p_z^2}{2}- p_y \, x +
    p_x \, y \\
    & + \frac{1}{1+ e \, \cos f} \left(  \frac{1}{2}\, e \, (x^2+y^2+z^2)
    \cos f  \right. \\
    & \left. -  \frac{\mu}{\sqrt{(x-(1-\mu))^2+y^2+z^2}} -
    \frac{1-\mu}{\sqrt{(x+\mu)^2+y^2+z^2}} \right)
  \end{aligned}
\end{equation}
where the independent variable, denoted by $f$, corresponds to the
true anomaly of the secondary body, the parameter $\mu\in
(0,\frac{1}{2}]$ denotes the reduced mass so that the masses of
  $P_1,P_2$ are $1-\mu,\mu$ respectively, and $e$ denotes the
  eccentricity of the elliptic motion.

The main advantage of using rotating--pulsating variables is that the
Hamilton equations of (\ref{eq:ori3bp}) have five equilibrium points
$L_1,\ldots ,L_5$ which are located in the same orbital positions
$(x_{L_i},y_{L_i},0)$ of the corresponding circular problem
characterized by the same value of $\mu$. We here focus our analysis
on the collinear equilibrium points $L_1,L_2$ identified by
$(x,y,z,p_{x},p_y,p_z)=(x_{L_i},0,0,0,x_{L_i},0)$.  For each selected
equilibrium $L_i$ it is convenient to introduce variables
$(\mathbf{q},\mathbf{p})=(q_1,q_2,q_3,p_1,p_2,p_3)$:
\begin{equation}\label{eq:expanxL1}
  \begin{aligned}
    x & = q_1 + x_{L_j}~, \quad     &p_x & = p_1~, \\
    y & = q_2~, \quad     &p_y & = p_2 + x_{L_i}~, \\
    z & = q_3~, \quad     &p_z & = p_3 \\
  \end{aligned}
\end{equation}
such that the equilibrium point $L_i$ is in the origin of the phase-space, 
and consider the Taylor expansion of $h$  in $(\mathbf{q},\mathbf{p})$:
\begin{equation}\label{eq:expandedH0}
H (\mathbf{q},\mathbf{p},f;e) = H_2 + H_3+ \ldots 
\end{equation}
where each term $H_j(\mathbf{q},\mathbf{p},f;e)$ is a polynomial 
of degree $j$ in the variables $(\mathbf{q},\mathbf{p})$. Notice that  
the zero-order term $H_0 (f;e)$ has been removed from the Hamiltonian, and 
the term of order $1$ vanishes because we are expanding the Hamiltonian at an equilibrium point.

Since paper~\cite{Conley67}, the dynamics originating at the
Lagrangian points $L_1,L_2$ of the CRTBP have been used to study the transits of
motions at low energies between the regions of the space which are
'internal' or 'external' with respect to the two massive bodies.  The
analysis is obtained from two properties of the CRTBP: the existence
of a global first integral, the so-called Jacobi integral, and of the
center, stable and unstable manifolds originating at the equilibria
$L_1,L_2$. In fact, close to a center-center-saddle equilibrium point
of a Hamiltonian system, provided that certain non-resonance
conditions are satisfied, the Hamiltonian has Birkhoff normal forms of
large order:
\begin{equation}\label{eq:BTnfcirc}
K (\mathbf{Q},\mathbf{P}) = K_2 (\mathbf{Q},\mathbf{P})+ K_4(\mathbf{Q},\mathbf{P})+ \ldots +K_N(\mathbf{Q},\mathbf{P})+R_{N+1}(\mathbf{Q},\mathbf{P})~,
\end{equation}
where $(\mathbf{Q},\mathbf{P})$ are canonical variables defined in a
neighbourhood of the equilibrium point\footnote{Note that in the
  variables $(\mathbf{Q},\mathbf{P})$ the couple $Q_3,P_3$ is no more
  identified with the vertical variables.  The planar problem is
  instead obtained for $Q_2,P_2=0$.}; each term
$K_j(\mathbf{Q},\mathbf{P})$ is an autonomous polynomial of degree $j$
in the variables $(\mathbf{Q},\mathbf{P})$ and is integrable, in the
sense that it depends on the variables only through the combinations
$(Q_1^2+P_1^2)/2$, $(Q_2^2+P_2^2)/2$ and $Q_3P_3$. $R_{N+1}
(\mathbf{Q},\mathbf{P},f;e)$ is the remainder of the Taylor expansion
of $K$, containing polynomials from order $N+1$. In~\cite{Conley67}
only the linear approximation and the planar problem were considered
(i.e. $N=2$ and $Q_3=P_3=0$ in (\ref{eq:BTnfcirc})), but later higher
non-linear normal forms were computed, as in
\cite{Simo99,JM99,GJMS,PY,GomKoLoMaMaRoss,masdemont05,KLMR06,giorgilli12,LG16,GL18,pucacco19,PG20},
etc.  In these papers the dynamics of transits is obtained by
approximating a Birkhoff normal form (\ref{eq:BTnfcirc}) of large
order $N$ with the integrable Hamiltonian (we refer to
\cite{sanders07,meyer09} for an introduction to polynomial normal
forms):
\begin{equation}\label{eq:BTnfint}
{\cal K} (\mathbf{Q},\mathbf{P}) = K_2 (\mathbf{Q},\mathbf{P})+ K_4(\mathbf{Q},\mathbf{P})+ \ldots +K_N(\mathbf{Q},\mathbf{P}) .
\end{equation}
From the values of the first integrals of Hamiltonian~\eqref{eq:BTnfint}, 
\begin{displaymath}
{\cal I}_1=  {Q_1^2+P_1^2\over 2}~~,
{\cal I}_2=  {Q_2^2+P_2^2\over 2}~~,
\,\,{\cal I}_3=Q_3P_3~~,
\end{displaymath}
one obtains a complete classification of the transit orbits according
to the approximate Hamiltonian (\ref{eq:BTnfint}). In fact, fixed
values of ${\cal I}_1,{\cal I}_2\geq 0$ and $Q_3,P_3=0$ define an
invariant manifold ${\cal M}_{{\cal I}_1,{\cal I}_2}$ for the flow
$\phi_{\cal K}$ of Hamiltonian ${\cal K}$ supporting quasi-periodic
motions, and the set $Q_3=0,P_3\ne 0$ (resp. $Q_3\ne 0,P_3=0$) defines
the local stable (resp. unstable) manifold of ${\cal M}_{{\cal
    I}_1,{\cal I}_2}$.  The values of ${\cal I}_3\ne 0$ identify the
transit properties: either the motions approach the Lagrange
equilibrium arriving from the direction of $P_1$ or $P_2$ and then
bounce back, either they transit from one side to the other one of the
equilibrium point. 
The stable and unstable manifolds of ${\cal M}_{{\cal I}_1,{\cal I}_2}$
are separatrices for the transit properties. Moreover, the conservation of
the Jacobi integral forces all the low energy transits to occur close
the Lagrangian points, due to the peculiar shape of the realms of
admissible motions, forming a bottleneck close to $L_1,L_2$.

This description of the dynamics close to $L_1,L_2$ is 
obtained by neglecting the remainder $R_{N+1}(\mathbf{Q},\mathbf{P})$, and 
therefore is affected by errors. By considering also the remainder 
$R_{N+1}$, the functions ${\cal I}_j$ may not be first 
integrals for the non approximated Hamiltonian $K$. In particular, while a 
center manifold survives the perturbation, its invariant tori could not, and a 
KAM theory should be implemented. Instead, the errors due to the remainder 
$R_{N+1}$ are less effective on the transits which occur in short  
time intervals, we refer to Section 2 for more details.
\vskip 0.4 cm In this paper we show how the previous discussion about
transit orbits extends to the elliptic restricted three--body problem,
despite the lack of a global first integral. In the ERTBP the
Lagrangian points remain equilibria of the non-autonomous Hamilton
equations of (\ref{eq:ori3bp}) and are therefore fixed points of the
Poincar\'e map defined by the Hamiltonian flow at time $2\pi$ of the
Hamilton equations of (\ref{eq:ori3bp}). Therefore, at least for small
values of the eccentricity, one may still follow the road to define
center, stable and unstable manifolds of the Poincar\'e map, providing
motions which librate close to the Lagrangian points, as well as
motions which are asymptotic to these librations. Unfortunately, these
motions would apparently be less useful for the characterization of
the transits since global first integrals are not known for the
ERTBP. Therefore, on the one hand we cannot define the realms of
motions which are forbidden or admissible for a certain value of the
Jacobi integral, thus making impossible a sharp definition of transit
motion; on the other hand we are not able to identify invariant
3-dimensional (5-dimensional for the spatial problem) level sets of
the phase--space disconnected by stable and unstable manifolds related
to the dynamics originating at $L_1,L_2$.

To overcome these issues we construct local normal forms for the
Hamiltonian (\ref{eq:ori3bp}), which we call Floquet-Birkhoff normal
forms, which are integrable and autonomous up to a suitable large
order $N$:
\begin{equation}\label{eq:BTnf0}
K (\mathbf{Q},\mathbf{P},f;e) = K_2 (\mathbf{Q},\mathbf{P};e)+ K_4(\mathbf{Q},\mathbf{P};e)+ \ldots +K_N(\mathbf{Q},\mathbf{P};e)+R_{N+1}(\mathbf{Q},\mathbf{P},
f;e)
\end{equation}
where $(\mathbf{Q},\mathbf{P})$ are canonical variables defined in a
neighbourhood ${\cal B}$ of the selected Lagrangian point $L_i$; each
term $K_j(\mathbf{Q},\mathbf{P};e)$ is an autonomous polynomial of
degree $j$ in the variables $(\mathbf{Q},\mathbf{P})$ and is
integrable in the sense that it depends on the
$(\mathbf{Q},\mathbf{P})$ only through the combinations
$(Q_1^2+P_1^2)/2$, $(Q_2^2+P_2^2)/2$ and $Q_3P_3$. The Taylor
expansion of the remainder $R_{N+1}(\mathbf{Q},\mathbf{P},f;e)$
contains polynomials from order $N+1$, and is possibly dependent
periodically on $f$. The first term of the expansion,
$$
K_2 (\mathbf{Q},\mathbf{P};e) = \sigma_1 {Q_1^2+P_1^2\over 2}+ \sigma_2 {Q_2^2+P_2^2\over 2}
+\lambda Q_3P_3 ~,
$$ represents the linear approximation of the system obtained from a
Floquet transformation defined at the Lagrange equilibrium and an
additional linear transformation which puts to evidence the
center-center-saddle nature of the equilibrium. A
combination of the Floquet theory and Birkhoff normalizations has been
used to study the stability of transversely elliptic periodic
orbits~\cite{M78,M05}, and for the equilateral equilibria
$L_4$, $L_5$ of the ERTBP~\cite{tesi}.  The development of the
Floquet-Birkhoff normal form~\eqref{eq:BTnf0} at the equilibria $L_1$,
$L_2$ meets additional complexity in the control of the numerical
precision and in the definition of the Floquet transformation. On one
hand, the partially hyperbolic nature of these equilibrium points may
be responsible of large errors in the computation of the coefficients
of the normal form.
On the other hand, we need to fix a gauge in the definition of the
Floquet transformation. As it is well know, the Floquet
transformation is not unique, since its definition depends on the arbitrary
choice of a logarithm of the monodromy matrix $\mathbf{\Phi_e}$,
computed at the Lagrange equilibrium. If one is interested only in the
dynamics of the Hamilton equations linearized at the equilibrium
point, as in the traditional Floquet theory, any choice of the
logarithm of $\mathbf{\Phi_e}$ can be used to define the Floquet
transformation. But, since our project is to improve the Floquet
approximation to higher orders, we need to select the logarithm of the
monodromy matrix providing a close to the identity Floquet
transformation. In fact, if the Floquet transformation is not close to the
identity, the Fourier expansion with respect to $f$ of the transformed
Hamiltonian contains a large number of terms with
large coefficients, so that the computation of the Birkhoff normal
forms saturates the computer memory at low normalization orders $N$.
All the technical details about the Floquet transformation that we
use and the Birkhoff transformations providing the normal form
(\ref{eq:BTnf0}) will be given in Section~\ref{sec:canfloquet}.

Finally,
we describe the transits which occur close to the Lagrangian points
using the dynamics of the approximated Hamiltonian:
\begin{equation}\label{eq:BTnfintell}
{\cal K} (\mathbf{Q},\mathbf{P};e) = K_2 (\mathbf{Q},\mathbf{P};e)+ K_4(\mathbf{Q},\mathbf{P};e)+ \ldots +K_N(\mathbf{Q},\mathbf{P};e)
\end{equation}
which is obtained by neglecting the small remainder 
$R_{N+1}(\mathbf{Q},\mathbf{P}f;e)$ in the Floquet-Birkhoff normal
form (\ref{eq:BTnf0}).  
\vskip 0.4 cm This paper is organized as
follows. Section~\ref{sec:overview} is dedicated to an overview of the
results about the dynamics of the ERTBP that follow from the
construction of the Floquet-Birkhoff normal forms (\ref{eq:BTnf0}).
In Section~\ref{sec:canfloquet} we provide all the analytic details of
the construction of the Floquet-Birkhoff normal forms. In
Section~\ref{sec:experiments} we show the application
of the method for the Earth-Moon Elliptic three-body problem. Finally,
we provide Conclusions.

\section{An overview of the results about the dynamics of the ERTBP 
following from the Floquet-Birkhoff normal forms}\label{sec:overview}

\begin{figure*}

\includegraphics[width=1\columnwidth]{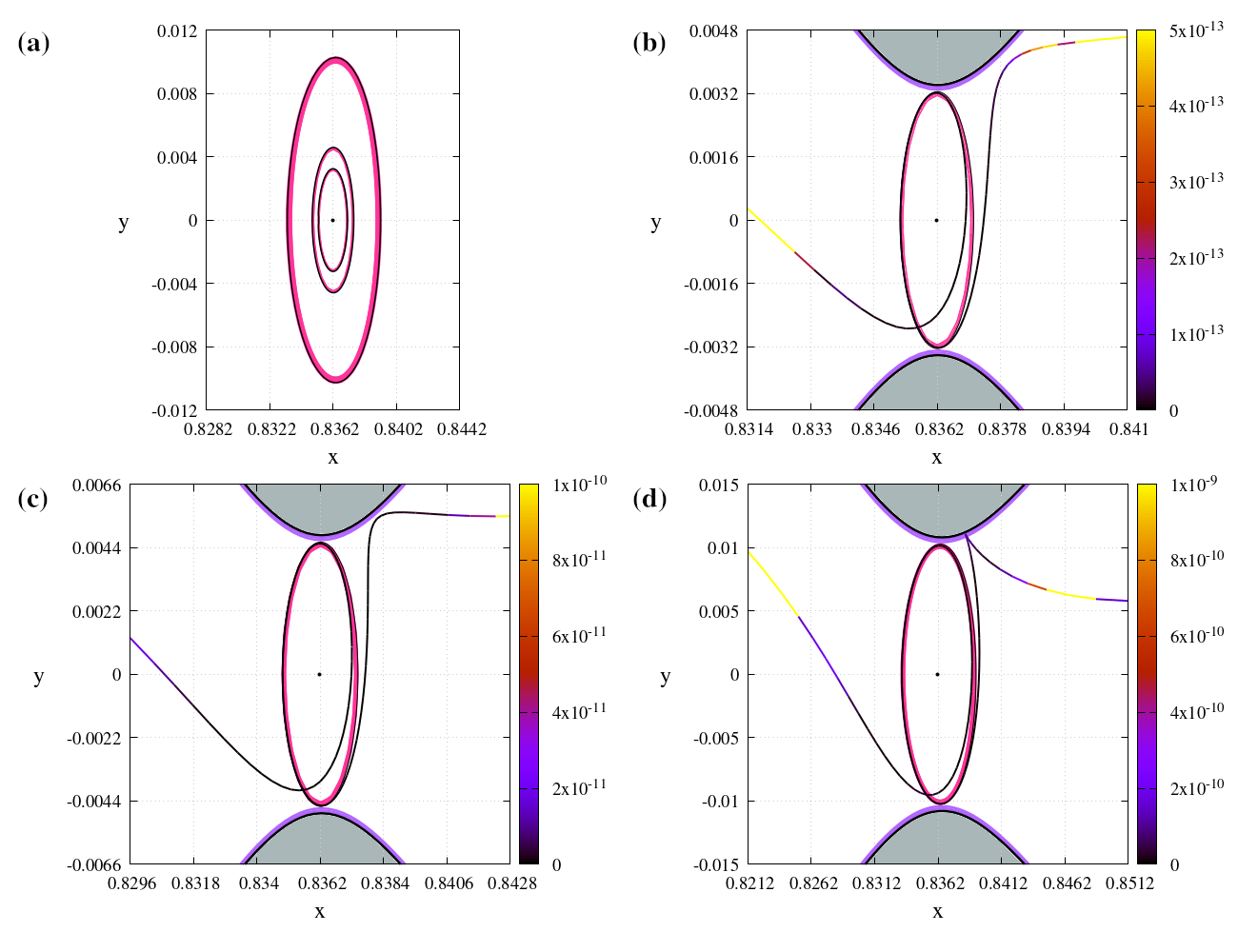}

\caption{\small Representation of sample orbits transiting at $L_1$
  for different values of the initial local energy $\kappa = $
  $2.336\snot[-5]$ (b), $4.672\snot[-5]$ (c), $2.336\snot[-4]$ (d),
  for the planar ERTBP defined by $\mu = 0.0123$, $e=0.0549006$
  (Earth-Moon ERTBP) and normalization order $N=8$.  Panel (a)
  displays in pink the projection on the Cartesian plane $xy$ of the
  sets ${\cal M}_{{\cal I}_1,0}$ corresponding to the three values of
  $\kappa$ (the section at $f=0$ for each energy is represented in
  black); panels (b), (c), (d) display examples of transiting orbits
  for the aforementioned energies.  The initial conditions providing
  the orbits have been found according to the Floquet-Birkhoff normal
  forms (more details will be given in Section~\ref{sec:experiments}),
  the orbits have been then obtained by numerically integrating the
  Hamilton equations of Hamiltonian (\ref{eq:ori3bp}).  The orbits are
  represented using a color scale which indicates the variation of the
  local energy with respect to the initial value, as the orbit
  proceeds. The 'zero velocities curves' for all $f$ are in the thin
  purple bands, delimited on the external side by the black curves;
  the vector $-\nabla {\cal V}_*$ points outward with respect to the
  shaded gray areas.}
  \label{fig:kappa1}
\end{figure*}

\begin{figure*}
\includegraphics[width=1\columnwidth]{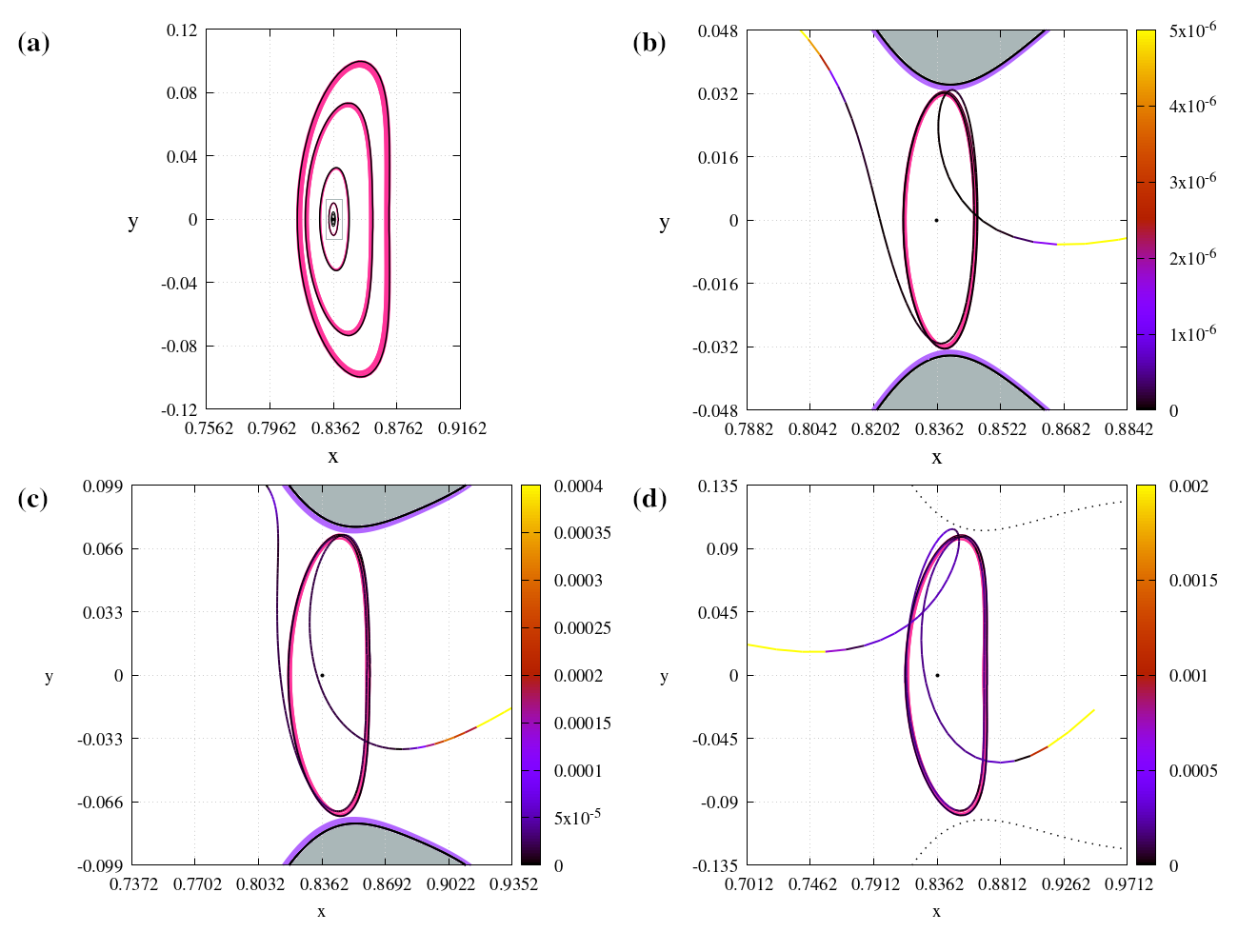}
  \caption{\small As for Fig. \ref{fig:kappa1} for the larger values
    of the local energies $\kappa = 2.329\snot[-3]$ (b); $\kappa =
    1.150 \snot[-2]$ (c) and $\kappa = 2.044\snot[-2]$ (d). To
    appreciate the larger amplitude of these orbits, we plot a small gray
    rectangle in panel (a) corresponding to panel (a) of Fig. 1. We do
    not report the zero velocity curves in panel (d), since for these
    large libration we expect a good conservation of the local energy
    only very close to the set ${\cal M}_{{\cal I}_1,0}$ (see Section
    4.2 for more details).}
  \label{fig:kappa2}
\end{figure*}

In this Section we present an overview of the key consequences
of the existence of Floquet-Birkhoff normal forms on the
dynamics of the ERTBP. We denote by ${\cal B}$ the 
neighbourhood of the origin in ${\Bbb R}^6$ where the Floquet-Birkhoff 
normal form is defined. It is not restrictive to assume 
that ${\cal B}$ has the form 
${\cal B}=B_1(\rho)\times B_2(\rho)\times B_3(\rho)$
where $B_j(\rho)$ is a two-dimensional Euclidean neighbourhood of 
$(Q_j,P_j)=(0,0)$ of radius $\rho$, and that the canonical transformation:
$$
\mathbf{(q,p)}=\Psi(\mathbf{Q},\mathbf{P},f;e)
$$
conjugating the Floquet-Birkhoff normal form (\ref{eq:BTnf0}) to
Hamiltonian (\ref{eq:expandedH0}) is well defined in for
all $\mathbf{(Q,P)}\in {\cal B}$ and all $f\in [0,2\pi]$. 
\vskip 0.4 cm
\noindent 
{\bf 1. Local energy.} Since 
Hamiltonian ${\cal K}$ in Eq.~(\ref{eq:BTnfintell}) is autonomous it is 
convenient to describe the motions defined by its Hamiltonian flow 
$\phi_{\cal K}$ for fixed values $\kappa$ of the function ${\cal K}$, which we 
call local energy.  In Figures
\ref{fig:kappa1} and \ref{fig:kappa2} we provide some examples of
 conservation of the local energy for solutions of the Earth--Moon ERTBP 
transiting in a neighbourhood of the Lagrangian point $L_1$. 
Since the Lie derivative of the local 
energy is proportional to the derivatives of the remainder $R_{N+1}$,
the variations of ${\cal K}$ in the flow of the complete Hamiltonian
are very small closer to the Lagrangian point, as in 
Figure \ref{fig:kappa1}, and become more important for the large librations
we reach in the examples of Figure  \ref{fig:kappa2}.    
\vskip 0.4 cm
\noindent
{\bf 2. Zero velocity surfaces.} By representing the Hamiltonian 
(\ref{eq:BTnfintell}) in the translated Cartesian 
variables $(\mathbf{q},\mathbf{p})$ we obtain a 
function $\hat {\cal K}(\mathbf{q},\mathbf{p};f)$ which is an approximate 
local first integral defined in a neighbourhood of the Lagrangian point $L_i$. 
For any value of $f$, and any small value $\kappa$ of the local energy,
we define the {\it zero velocity surfaces} through the equation:
\begin{equation}\label{zvs}
\hat {\cal K}(q_1,q_2,q_3,-q_2,q_1,0,f;e)=\kappa  ,
\end{equation}
and the {\it zero velocity curves} for the planar problem. In the ERTBP
the zero velocity surfaces do not strictly provide a barrier for the motions 
as for the surfaces obtained from the Jacobi constant for the CRTBP. 
Nevertheless, since when $\dot {\mathbf{q}}= (0,0,0)$ we have
$\ddot {\mathbf{q}} = -\nabla {\cal V}_*(\mathbf{q},f;e)$ with
{\small 
\begin{equation}\label{V*}
\begin{aligned}
{\cal V}_*& (\mathbf{q},f;e) = \left (-{1\over 2}(x^2+y^2)+
\frac{1}{1+ e \, \cos f} \left( \frac{1}{2}\, e \, (x^2+y^2+z^2) \cos
f \right.\right. \\ & \left.\left. -
\frac{\mu}{\sqrt{(x-(1-\mu))^2+y^2+z^2}} -
\frac{1-\mu}{\sqrt{(x+\mu)^2+y^2+z^2}} \right)\right )_{x=
  q_1+x_{L_1},y=q_2,z=q_3}~,
 \end{aligned}
\end{equation}}
the direction of the vector $ -\nabla {\cal V}_*(\mathbf{q},f;e)$ at the
zero velocity surfaces provide an indication of the repelling effect of
the surfaces on the motions (the position of $\mathbf{q}(f)$
should be compared with the surface defined by the value $f$ of the true
anomaly).  A numerical computation of the zero-velocity 
curves is reported 
in Figures~\ref{fig:kappa1},~\ref{fig:kappa2}, where for all the cases
the vector  $-\nabla {\cal V}_*$ points outward with respect to the 
shaded area.
\vskip 0.4 cm
\noindent
{\bf 3. Transit orbits.} For $\kappa > 0$ the level set:  
\begin{displaymath}
{\cal M}_{\kappa}= \{  (\mathbf{Q},\mathbf{P})\in {\cal B}:\ \ 
{\cal K} (\mathbf{Q},\mathbf{P};e)=\kappa\ \ ,\ \ Q_3,P_3=0\}
\end{displaymath}
contains a collection of sets ${\cal M}_{{\cal I}_1,{\cal I}_2}$
invariant for the approximated flow $\phi_{\cal K}$. Its local stable
and unstable manifolds $W^{s,loc}_\kappa,W^{u,loc}_\kappa$ are
defined by $Q_3=0,P_3\ne 0$ or $Q_3\ne 0,P_3= 0$
respectively. The local stable and unstable manifolds are
separatrices for the transits occurring close to the Lagrangian point,
and we recover from the circular case the classification of transit
motions according to the position of the orbits expressed in the
variables $Q,P$ with respect to $W^{s,loc}_\kappa,W^{u,loc}_\kappa$.

A transit is defined as a motion $(\mathbf{Q}(f),\mathbf{P}(f))$ of
$\phi_{\cal K}$ in the interval of $f\in [f_0,f_1]$ such that:
$(\mathbf{Q}(f),\mathbf{P}(f)) \in {\cal B}$ for all $f\in (f_0,f_1)$,
and both endpoints
$(\mathbf{Q}(f_0),\mathbf{P}(f_0)),(\mathbf{Q}(f_1),\mathbf{P}(f_1))$
belong to the border of ${\cal B}$. Moreover, we have ${\cal I}_3>0$
along the motion. In fact, since in the approximated flow $\phi_{\cal K}$ the
variables $Q_1,P_1,Q_2,P_2$ oscillate periodically, the motion
$(\mathbf{q}(f),\mathbf{p}(f))$ is the superposition of quasi-periodic
oscillations and of an hyperbolic motion which defines the transit
property. For small values of the eccentricity $e$, the Cartesian
variables $q_1,q_2$ are related to the variables
$\mathbf{(Q,P)}$ by
\begin{eqnarray}\label{xy2QP}
q_1 &=& a (Q_3-P_3) +b P_1 +.....  \cr
q_2 &=& c (Q_3+P_3) +d Q_1 + ..... 
\end{eqnarray}
where $a,b,c,d$ are numbers depending on $\mu$, and the 'dots'
indicate both linear contributions which are not present when $e=0$,
or non-linear contributions. Therefore the transits
from/to negative to/from positive values of $q_1$ occur for ${\cal
  I}_3=Q_3P_3>0$.

\begin{figure*}
  \centering
  \includegraphics[width=1.\columnwidth]{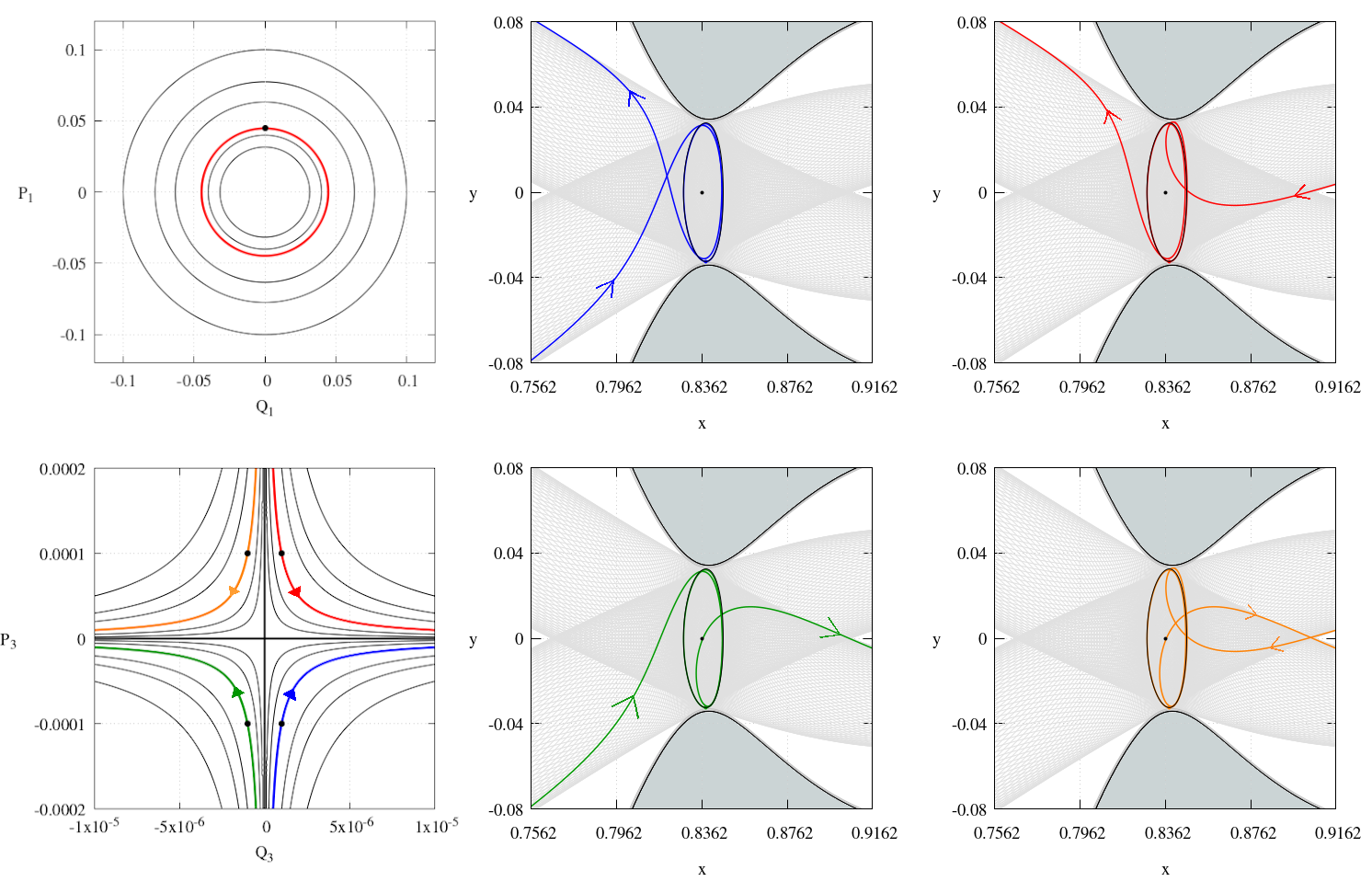}
  \caption{\small Transit and non-transit orbits numerically computed
    from initial conditions chosen according to different values of
    the normalized variables $Q_3,P_3$.  The initial conditions are:
    $Q_1=0$, $P_1= 1/(10\,\sqrt{5})$, $Q_2,P_2=0$ and $f = 0$ for all
    the orbits, while $Q_3 = 1\snot[-6]$, $P_3 = -1\snot[-4]$ for the
    blue orbit, $Q_3 =1\snot[-6]$, $P_3 = 1\snot[-4]$ for the red orbit,
    $Q_3 = -1\snot[-6]$, $P_3 = -1\snot[-4]$ for the green orbit and
    $Q_3 = -1\snot[-6]$, $P_3 =1\snot[-4]$ for the orange orbit. The center
    and right panels show the projection of the orbits in the original
    $xy$ Cartesian variables. The left (upper and lower) panels show
    the projection of the orbits in normalized variable planes
    $(Q_1,P_1)$ (where the four orbits overlap) and $(Q_3,P_3)$
    respectively. The black points are the initial conditions of the
    orbits. The black elongated curve corresponds to the section $f=0$
    of the 2-d torus ${\cal M}_{{\cal I}_1,0}$; the gray
    orbits are in the stable and unstable manifolds of ${\cal
      M}_{{\cal I}_1,0}$.  The black-bold curves are the zero velocity
    curves for $f =0$.  The arrows indicate the direction of the
    motion for increasing values of $f$.  }
  \label{fig:tubos}
\end{figure*}

\begin{figure*}
  \centering
  \includegraphics[width=1.\columnwidth]{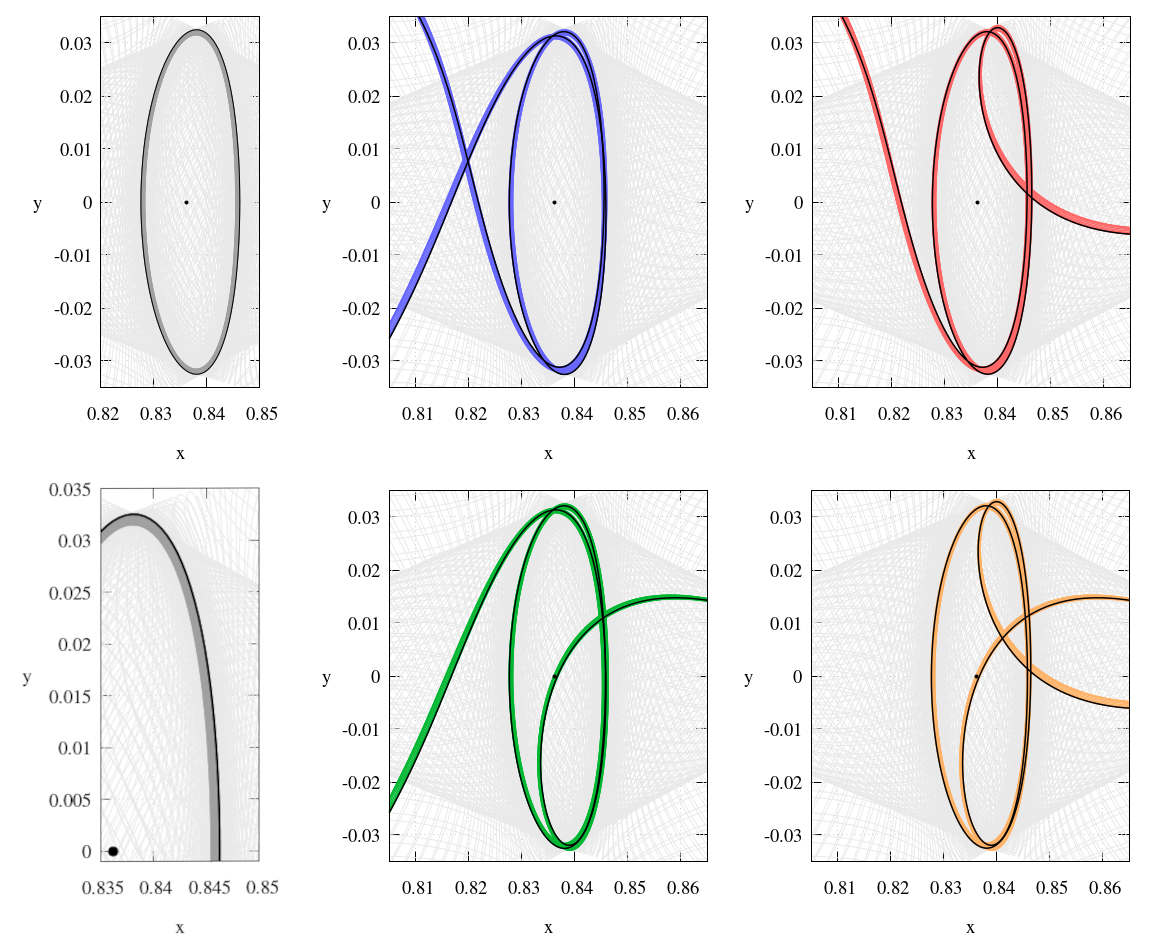}
  \caption{\small As Figure~\ref{fig:tubos}, but considering a set of
    15 different initial values of $f$, instead of a single one,
    represented as a color band. The amplitude of the bands
    indicates the influence of the non
    null eccentricity. In all panels, the
    black orbit corresponds to the orbits show in the examples of
    Fig. \ref{fig:tubos}.}
  \label{fig:tubosconnec_new}
\end{figure*}

In Fig.~\ref{fig:tubos} we represent four transit and non-transit
planar orbits whose initial conditions have been chosen using the
Floquet-Birkhoff normal form, as well as their projections on the
planes of the normalized variables $(Q_1,P_1)$ and $(Q_3,P_3)$. The
red and green orbits are of transit type, while the blue and orange
orbits are non-transit type. In Figure~\ref{fig:tubosconnec_new} we
represent a family of transit and non--transit orbits obtained from
the same initial conditions for the variables $\mathbf{Q},\mathbf{P}$
used for Fig.~\ref{fig:tubos}, but different initial values of $f$,
thus showing the effect of the non null eccentricity. Transit orbits
in the spatial case are presented in Section~\ref{sec:transits}.

\vskip 0.4 cm
\noindent
{\bf 4. Error estimates, local diffusion, Arnold diffusion.} By neglecting the small remainder $R_{N+1}(\mathbf{Q},\mathbf{P}, f;e)$ we are introducing errors
which are mainly due to the fact that the functions ${\cal I}_j$ may not be
first integrals. Nevertheless, in a neighbourhood of radius $\varrho$ of the equilibrium, their Lie derivatives are small,
$$
{d\over df}{\cal I}_j = \{ {\cal I}_j , R_{N+1}\}=
{\cal O}(\varrho^{N+1})\ \ \ \ j=1,2,3 ~,
$$
and therefore their cumulative variation
\begin{equation}\label{varIj}
|{\cal I}_j(f_1)-{\cal I}_j(f_0)| \leq
\norm{f_1-f_0}\sup_{(Q,P)\in {\cal B},f \in [f_0,f_1]} \norm{\{ {\cal
    I}_j , R_{N+1}\}}
\end{equation}
is small when the transits occur in a small interval $[f_0,f_1]$.
The order of normalization $N$ fixes a lower bound on ${\cal I}_3>0$, such that
the transits occurring with larger values of ${\cal I}_3$ are well
approximated by the transits obtained from the integrable Hamiltonian
${\cal K}$.

In fact, let us consider the following argument, which is here presented in an
heuristic way. By assuming for simplicity ${\cal K}=K_2$, an orbit with ${\cal I}_3>0$ entering to the set ${\cal B}$ at $f_0$ will exit from ${\cal B}$
after a time interval $\Delta f$ of order:
\begin{displaymath}
\Delta f \sim {1\over \lambda}\ln {\rho^2\over {\cal I}_3}   .
\end{displaymath}
For each given normalization order $N$
there exists a constant $C$ (uniform with respect to the choice
of the initial conditions) such that in the same interval the flow of the
ERTBP will change the value of ${\cal I}_3$ no more than
\begin{displaymath}
\Delta {\cal I}_3 \leq C \Delta f \rho^{N+1} \sim {C\over \lambda}\rho^{N+1}
\ln {\rho^2\over {\cal I}_3} .
\end{displaymath}
Therefore, for any chosen small $\epsilon$, during the transit
the variation of ${\cal I}_3$ is smaller than $\epsilon\, {\cal I}_3(f_0)$ as soon as
\begin{displaymath}\label{lowerbound}
 {{\cal I}_3(f_0) \over \ln {\rho^2\over {\cal I}_3(f_0)}} \geq {C \over \epsilon \lambda}
\ \rho^{N+1}  ,
\end{displaymath}
providing a lower threshold for the value of ${\cal I}_3$, depending
on the normalization order $N$. In the time interval $\Delta f$ the
actions ${\cal I}_1,{\cal I}_2$ (as well as the local
energy $\kappa$) have small variations according to inequality \eqref{varIj}.

Instead, for motions which remain close to $Q_3,P_3=0$ for long times,
our methods do not allow to rule out the accumulation of errors
providing a possible very slow local diffusion of ${\cal I}_1,{\cal
  I}_2$, driving the motion outside the set ${\cal B}$.

Finally, while each individual transit occurring with values of ${\cal
  I}_3$ satisfying (\ref{lowerbound}) preserves (approximately) the
value of the local energy $\kappa$ up to a very small variation,
orbits of the ERTBP which exit the set ${\cal B}$ are allowed to
re-enter the set ${\cal B}$ at a later time with a different value of
$\kappa$. A similar phenomenon, even if not related to transits but to
the homoclinic returns to the center manifold, has been proved for the
ERTBP using techniques of Arnold diffusion~\cite{CGdlL16}.


\section{Floquet-Birkhoff normalization of the Hamiltonian}\label{sec:canfloquet}

Let us consider the variables
$(\mathbf{q},\mathbf{p})=(q_1,q_2,q_3,p_1,p_2,p_3)$ introduced in
Eq.~\eqref{eq:expanxL1} and consider the Taylor expansion of $h$ in
$(\mathbf{q},\mathbf{p})$:
\begin{equation}\label{eq:expandedH}
H (\mathbf{q},\mathbf{p},f;e) = H_2 + H_3+ \ldots~,
\end{equation}
where each term $H_j(\mathbf{q},\mathbf{p},f;e)$ is a polynomial of
degree $j$ in the variables $(\mathbf{q},\mathbf{p})$ (notice that the
zero-order term $H_0 (f;e)$ has been removed from the Hamiltonian and
that the term of first order vanishes because we are expanding the
Hamiltonian at an equilibrium point). The term of second order is
{\small
\begin{equation}\label{eq:H2}
    H_2 (\mathbf{q},\mathbf{p},f;e)= \, \frac{p_1^2}{2} + \frac{p_2^2}{2} +
    \frac{p_3^2}{2}  - p_2 \, q_1 + p_1 \, q_2 
     + \, \frac{\beta\, (-2 q_1^2+q_2^2+q_3^2)}{1
    + e \cos f}  + \frac{e \cos f\, (q_1^2+q_2^2+q_3^2)}{2\,(1 + e \cos f)}
\end{equation}}
with
\begin{equation}\label{eq:beta}
  \beta = \frac{1}{2} \left(\frac{\mu}{\norm{1-x_{L_i} -\mu}^3} + \frac{1-\mu}{
\norm{x_{L_i}+\mu}^3} \right)~.
\end{equation}
We here use a combination of the Floquet theory and Birkhoff normalizations
to conjugate the Hamiltonian (\ref{eq:expandedH}) to a normal
form which is integrable and autonomous up to a suitable large order $N$:
\begin{equation}\label{eq:BTnf}
K (\mathbf{Q},\mathbf{P},f;e) = K_2 (\mathbf{Q},\mathbf{P};e)+ K_4(\mathbf{Q},\mathbf{P};e)+ \ldots +K_N(\mathbf{Q},\mathbf{P};e)+R_{N+1}(\mathbf{Q},\mathbf{P},
f;e)~.
\end{equation}
Each term $K_j(\mathbf{Q},\mathbf{P};e)$ is an autonomous
polynomial of degree $j$ in the variables $(\mathbf{Q},\mathbf{P})$
and is integrable, in the sense that it depends on the variables only
through the combinations $(Q_1^2+P_1^2)/2$, $(Q_2^2+P_2^2)/2$ and
$Q_3P_3$. The remainder $R_{N+1} (\mathbf{Q},\mathbf{P},f;e)$ of
the Taylor expansion of $K$ contains monomials from order $N+1$
and is possibly dependent on $f$. 
The Floquet-Birkhoff normal form is obtained from the
composition of:
\begin{itemize}
\item[(i)] a canonical Floquet transformation:
$$
(\mathbf{q},\mathbf{p})={\cal C}(f;e)(\mathbf{\tilde q},\mathbf{\tilde p})
$$
conjugating the Hamiltonian (\ref{eq:expandedH}) to an Hamiltonian:
\begin{equation}\label{eq:expandedHt}
\tilde H (\mathbf{\tilde q},\mathbf{\tilde p},f;e) = \tilde H_2 (\mathbf{\tilde q},\mathbf{\tilde p};e)+ \tilde  H_3(\mathbf{\tilde q},\mathbf{\tilde p},f;e)+ \ldots 
\end{equation}
where each term $ \tilde H_j(\mathbf{\tilde q},\mathbf{\tilde p},f;e)$
is polynomial of degree $j$ in the variables $\mathbf{\tilde
  q},\mathbf{\tilde p}$ and periodic in $f$ with period $2\pi$, while
$\tilde H_2 (\mathbf{\tilde q},\mathbf{\tilde p};e)$ is autonomous.

\item[(ii)] a linear canonical transformation:
\begin{equation}\label{eq:linD2}
 ( \mathbf{\tilde q}, \mathbf{\tilde p}) = {\cal D}(\mathbf{\hat Q}, \mathbf{\hat P})
\end{equation}
giving $\tilde H_2 (\mathbf{\tilde q},\mathbf{\tilde p};e)$ the normal form:
\begin{equation}\label{eq:k2}
k_2( \mathbf{\hat Q},\mathbf{\hat P})= \sigma_1{ {\hat Q}_1^2+{\hat
    P}_1^2\over 2}+\sigma_2{ {\hat Q}_2^2+{\hat P}_2^2\over 2}+\lambda
\hat Q_3 \hat P_3 .
\end{equation}
We denote by $k_j(\mathbf{\hat Q},\mathbf{\hat P},f;e)$ the image of all the 
other polynomials $k_j(\mathbf{\hat Q},\mathbf{\hat P};,f;e)=
\tilde H_j ( {\cal D}(\mathbf{\hat Q}, \mathbf{\hat P}),f;e)$.

\item[(iii)] a sequence of $N-2$ Birkhoff transformations giving the 
Hamiltonian the final normal form (\ref{eq:BTnf}), which we call
Floquet-Birkhoff normal form of order $N$. 
\end{itemize} 
\vskip 0.4 cm
Particular attention must be devoted to the construction of the
canonical Floquet transformation (i), whose definition is not unique
since it relies on the choice of a logarithm of the monodromy matrix
$\mathbf{\Phi_e}$ of the Hamiltonian flow of $H_2$. If one is
interested in the dynamics of the linearized Hamiltonian $H_2$, any
choice of the logarithm of $\mathbf{\Phi_e}$ can be used to define the
Floquet transformation. But, since our project is to make autonomous
also polynomials of higher order, we need to select the logarithm of
the monodromy matrix providing a close to the identity Floquet
transformation
$$
{\cal C}(f;e) ={\mathbb I}+e C(f;e)~;
$$ 
more details will be given in the Subsections below. 

The linear transformation (ii) exists provided the monodromy matrix
$\mathbf{\Phi_e}$ has a couple of
real eigenvalues $e^{2\pi\lambda},e^{-2 \pi\lambda}\ne 1$ and two couples of  
complex conjugate eigenvalues $a_1\pm i b_1$, $a_2\pm i b_2$ with
$a_1^2+b_1^2=a_2^2+b_2^2=1$, as it happens for $e=0$. 

Finally, the Birkhoff transformations (iii) exist provided the frequencies
$\sigma_1,\sigma_2$ appearing in (\ref{eq:k2}) and the frequency 
$\sigma_3=1$ associated to the motion of the primaries 
have no resonances of order smaller or equal than $N$: 
\begin{equation}\label{eq:res}
j_1 \sigma_1+ j_2\sigma_2 +j_3 \ne 0 \ \ \ \ \forall 
(j_1,j_2)\in {\Bbb Z}^2: \,\, \norm{j_1}+\norm{j_2}\in [1,N] , \quad \forall
j_3\in {\Bbb Z}\ \  .
\end{equation}
Our results improve the Floquet theory for those values of $\mu,e$
such that the frequencies $\sigma_1,\sigma_2$ have no resonances of
order smaller or equal than $N=3$.  To give an idea of the possible
resonances occurring for the lowest values of $N$, in
Table~\ref{tab:res} we provide the computation of the resonances for
$e = 0$.


\begin{table}[h]
  \centering
\begin{tabular}{| c | c | c | c | c |}
  \hline
  \hline
  $\phantom{L_i}$ & $j_1$ & $j_2$ & $j_3$ & $\mu$ \\
  \hline
  $L_1$ & -1 & 2 & -2 & $2.70101\snot[-4]$\\
  \hline
  $L_2$ & -1 & 0 & -2 & $4.00200\snot[-4]$\\
  $L_2$ & -2 & 0 &  3 & $2.59916\snot[-1]$\\
  $L_2$ & -2 & 1 &  2 & $3.88166\snot[-3]$\\
  $L_2$ & -1 & -1 & 3 & $2.12951\snot[-1]$\\
  $L_2$ & 0 & 2 & -3 & $ 1.70749\snot[-1]$\\
  \hline
\end{tabular}
\caption{Values of $\mu$ for which lower order resonances
  ($N=3$) occur in the vicinity of $L_1$ or $L_2$, see Eq.~\eqref{eq:res}.}
\label{tab:res}
\end{table}

\subsection{The canonical Floquet Transformation}\label{sub:tradi}

The Floquet theorem~\cite{Floquet} provides a representation of the solutions of
periodic linear differential equations of the form
\begin{equation}\label{eq:flo1}
\dot{\mathbf{x}} = A(t) \, \mathbf{x}
\end{equation}
where $\mathbf{x} \in \mathbb{R}^{n}$ and the matrix $A(t)$ is a
regular function of period $T$.  The principal fundamental matrix for
the system~\eqref{eq:flo1} is the function
$\Phi(t)$ whose columns
are $n$ linearly independent solutions of~\eqref{eq:flo1} such that
$\Phi(0) = \mathbb{I}$. The matrix $\mathbf{\Phi}:=\Phi(T)$ is usually
known as the monodromy matrix and its eigenvalues are the characteristic
multipliers of the system. The theorem states that $\Phi(t)$ can be written as
\begin{equation}\label{eq:flo2}
  \Phi(t) = {\cal C}(t) \, \mathrm{e}^{B \, t} \qquad \forall \, t \in \mathbb{R}
\end{equation}
where ${\cal C}$ is a non-singular 
$2T$-periodic function with ${\cal C}(0)=\mathbb{I}$ 
and $B$ is a real matrix satisfying
\begin{equation}\label{eq:flo3}
  \mathrm{e}^{2TB} = \mathbf{\Phi}^2~.
\end{equation}
When the matrix $A(t)$ is Hamiltonian, the matrix ${\cal C}(t)$ can 
be defined symplectic (see \cite{Cfloquet}). A key consequence of the theorem 
is the existence of a time-dependent change of coordinates:
\begin{equation}\label{eq:flo4}
  \mathbf{x} = {\cal C}(t) \, \mathbf{y}~,
\end{equation}
conjugating the linear system~\eqref{eq:flo1} to the autonomous one:
\begin{equation}\label{eq:flo5}
\dot{\mathbf{y}} = B \, \mathbf{y}~,
\end{equation}
with $B$ satisfying~\eqref{eq:flo3}. For our purposes we need to
define a canonical Floquet transformation which conjugates $H_2$ to an
autonomous quadratic Hamiltonian.  For a different application of the
Floquet theory, regarding the stability/instability transition of
the normal modes in the circular problem, see~\cite{CCP16}.

We denote by $\mathbf{x} =(\mathbf{q},\mathbf{p})$, 
$\mathbf{y} =(\mathbf{\tilde q},\mathbf{\tilde p})$ the
phase--space vectors in $\mathbb{R}^6$, and define
the matrix $A(f;e)$ by
\begin{equation}\label{eq:laM}
A(f;e) =  \mathbb{E} \nabla H_2  = 
  \left(
  \begin{array}{c c c c c c}
    0 &  1 & 0 & 1 & 0 & 0 \\
    -1 &  0 & 0 & 0 & 1 & 0 \\
    0 &  0 & 0 & 0 & 0 & 1 \\
    \frac{4 \beta - e \cos f}{1+ e \cos f} & 0 & 0 & 0 & 1 & 0 \\
    0 & - \frac{2 \beta + e \cos f}{1 + e \cos f} & 0 & -1 & 0 & 0 \\
    0  & 0  & - \frac{2 \beta + e \cos f}{1 + e \cos f} & 0 & 0 & 0 \\    
  \end{array}
  \right)  ,
\end{equation}
where $\mathbb{E}$ is the standard symplectic matrix 
of ${\mathbb R}^{6}$.  We denote by $\Phi(f;e)$ the 
principal fundamental matrix solution for the Hamilton
equations of $H_2$, and by ${\mathbf \Phi}_e :=\Phi(2\pi;e)$ the 
monodromy matrix. We have the following algebraic lemma.
\vskip 0.4 cm
\noindent 
{\bf Lemma.} {\it Assume that the matrix $\mathbf{\Phi}_e$ has a
  couple of real eigenvalues $e^{2\pi\lambda}>e^{-2 \pi\lambda}$ and
  two different couples of complex conjugate eigenvalues $a_1\pm i
  b_1$, $a_2\pm i b_2$ with $a_1^2+b_1^2=a_2^2+b_2^2=1$ and
  $b_1,b_2>0$. Then there exists a real symplectic matrix $C$, which
  is explicit function of the eigenvectors of $\mathbf{\Phi_e}$, such
  that the matrix defined by
 \begin{equation}\label{eq:laB}
\hat B_e(k_1,k_2)=  C  \left(
  \begin{array}{c c c c c c}
    0 &  0 & 0 & \pm(\omega_1+k_1) & 0 & 0 \\
    0 &  0 & 0 &   0 & \pm(\omega_2+k_2) & 0 \\
    0 &  0 & \pm\lambda & 0 & 0 & 0 \\
    \mp (\omega_1+k_1) & 0 & 0 & 0 & 0 & 0 \\
    0 &  \mp (\omega_2+k_2) & 0 & 0 & 0 & 0 \\
    0 & 0  & 0 & 0 & 0 & \mp\lambda \\    
  \end{array}
  \right)   
C^{-1}~,
\end{equation}
where
\begin{equation}\label{eq:omg}
\omega_j = {1\over 2\pi} \arccos(a_j)\ \ ,\ \ j=1,2
\end{equation}
$k_1,k_2$ are arbitrary integer numbers and the choice of 
the signs in the matrix depends on the eigenvectors of $\mathbf{\Phi_e}$, 
satisfies
\begin{equation}\label{logmonod}
  \mathrm{e}^{2\pi \hat B_e(k_1,k_2)} = \mathbf{\Phi}_e ~.
\end{equation}
The matrix $\hat B_e(k_1,k_2)$
is Hamiltonian, i.e.
\begin{equation}\label{hamb}
(\mathbb{E}\hat B_e(k_1,k_2))^T=\mathbb{E}\hat B_e(k_1,k_2)~,
\end{equation}
and defines a canonical Floquet transformation
\begin{equation}\label{floquet1}
 {\cal C}(f;e)= \Phi(f;e) \mathrm{e}^{-f \hat B_e(k_1,k_2)}
\end{equation}
which for $e=0$ is the identity matrix ${\cal C}(f;0) ={\mathbb I}$ if 
$k_1,k_2$ satisfy 
\begin{equation}\label{Bok1k2}
\hat B_0(k_1,k_2) = A_0\ \ ,\ \ A_0:=A(f;0)  .
\end{equation}
}

\begin{table}
\begin{tabular}{| c | c | c | c | c | c | c | c | c | c | c |}
  \hline
  \hline
  \phantom{I} & \multicolumn{5}{|c|}{$L_1$} & \multicolumn{5}{|c|}{$L_2$} \\
  \hline
  $\mu$ & $\lambda$ & $\Omega_1$ & $\Omega_2$ & $k_1$ & $k_2$ & $\lambda$ & $\Omega_1$ & $\Omega_2$ & $k_1$ & $k_2$ \\  
  \hline
$  1\snot[-6]$   & $2.5251$ & $	2.0818 $ & $2.0105 $ & $2  $ & $2$  & $2.4917$ & $2.0615 $ & $1.9897$ & $2  $ & $-2$\\
$  5\snot[-6]$   & $2.5371$ & $	2.0892 $ & $2.0180 $ & $2  $ & $2$  & $2.4800$ & $2.0544 $ & $1.9824$ & $2  $ & $-2$\\
$  1\snot[-5]   $   & $2.5447$ & $	2.0938 $ & $2.0227 $ & $2  $ & $2$  & $2.4728$ & $2.0500 $ & $1.9779$ & $2  $ & $-2$\\
$  5\snot[-5]    $   & $2.5710$ & $	2.1099 $ & $2.0392 $ & $2  $ & $2$  & $2.4481$ & $2.0350 $ & $1.9626$ & $2  $ & $-2$\\
$  0.0001    $   & $2.5877$ & $	2.1202 $ & $2.0497 $ & $2  $ & $2$  & $2.4328$ & $2.0258 $ & $1.9531$ & $2  $ & $-2$\\
$  0.0003    $   & $2.6241$ & $	2.1425 $ & $2.0726 $ & $2  $ & $2$  & $2.4005$ & $\bf 2.0063 $ & $1.9332$ & $\bf 2  $ & $-2$\\
$  0.0005    $   & $2.6464$ & $	2.1563 $ & $2.0867 $ & $2  $ & $2$  & $2.3812$ & $\bf 1.9947 $ & $1.9213$ & $\bf -2 $ & $-2$\\
$  0.0008    $   & $2.6709$ & $	2.1714 $ & $2.1021 $ & $2  $ & $2$  & $2.3606$ & $1.9823 $ & $1.9086$ & $-2 $ & $-2$\\
$  0.001     $   & $2.6840$ & $	2.1795 $ & $2.1104 $ & $2  $ & $2$  & $2.3497$ & $1.9758 $ & $1.9019$ & $-2 $ & $-2$\\
$  0.005     $   & $2.8176$ & $	2.2625 $ & $2.1954 $ & $2  $ & $2$  & $2.2441$ & $1.9129 $ & $1.8376$ & $-2 $ & $-2$\\
$  0.01	     $   & $2.9037$ & $	2.3166 $ & $2.2506 $ & $2  $ & $2$  & $2.1796$ & $1.8749 $ & $1.7987$ & $-2 $ & $-2$\\
$  0.03	     $   & $3.0917$ & $\bf 2.4355 $ & $2.3721 $ & $\bf 2  $ & $2$  & $2.0417$ & $1.7948 $ & $1.7168$ & $-2 $ & $-2$\\
$  0.05	     $   & $3.2054$ & $\bf 2.5081 $ & $\bf 2.4462 $ & $\bf -3 $ & $\bf 2$  & $1.9568$ & $1.7462 $ & $1.6673$ & $-2 $ & $-2$\\
$  0.08	     $   & $3.3258$ & $	2.5855 $ & $\bf 2.5251 $ & $-3 $ & $\bf -3$ & $1.8618$ & $1.6927 $ & $1.6128$ & $-2 $ & $-2$\\
$  0.1	     $   & $3.3879$ & $	2.6256 $ & $2.5660 $ & $-3 $ & $-3$ & $1.8095$ & $1.6635 $ & $\bf 1.5833$ & $-2 $ & $\bf -2$\\
$  0.2	     $   & $3.5927$ & $	2.7585 $ & $2.7015 $ & $-3 $ & $-3$ & $1.6048$ & $\bf 1.5526 $ & $\bf 1.4713$ & $\bf -2 $ & $\bf 1$\\ 
$  0.3	     $   & $3.7053$ & $	2.8321 $ & $2.7764 $ & $-3 $ & $-3$ & $1.4419$ & $\bf 1.4680 $ & $1.3871$ & $\bf 1 $ & $1$\\
$  0.49	     $   & $3.7832$ & $	2.8832 $ & $2.8283 $ & $-3 $ & $-3$ & $1.1696$ & $1.2589 $ & $1.2589$ & $1 $ & $1$\\
  \hline
  \hline
\end{tabular}
\caption{Values for $k_1$ (left) and $k_2$ (right) defined from
  Eq. (\ref{analytick1k2}), according to the values of
  $\Omega_1,\Omega_2$ numerically computed at both Lagrangian points
  $L_1,L_2$ for different values of $\mu$. The transitions in the
  values of $k_1,k_2$, occurring when $\Omega_1,\Omega_2$ cross half
  integer values, are highlighted in bold.}
\label{tablek1k2}
\end{table}

\vskip 0.4 cm
\noindent
{\bf Choice of the Logarithm of the monodromy matrix.} According to
Eq.~(\ref{logmonod}) the matrix $2\pi \hat B_e(k_1,k_2)$ is a
logarithm of the monodromy matrix $\mathbf{\Phi}_e$ for all the
choices of the integers $k_1,k_2$. For $k_1,k_2=0$ the matrix $2\pi
\hat B_e(k_1,k_2)$ is the principal logarithm, which we do not
identify as the more convenient choice. Precisely, we use 
Eq.~(\ref{Bok1k2}) to fix the values of $k_1,k_2$ in order to obtain a
close to the identity Floquet transformation. By denoting with $\pm
i\Omega_1,\pm i \Omega_2$ the complex eigenvalues of $A_0$ for a given
choice of $\mu$, the values of $k_1,k_2$ providing ${\cal C}(f;0)
={\mathbb I}$ are:
\begin{equation}\label{analytick1k2}
k_j = s_j\Omega_j - {1\over 2\pi}\arccos \left [\cos (2\pi \Omega_j)  
\right ]
\end{equation}
where $s_j=1$ if $\mod(\Omega_j,1)\in (0,1/2)$ while $s_j=-1$ if
$\mod(\Omega_j,1)\in (1/2,1)$. Therefore we may have bifurcations
when, by changing the parameter $\mu$, the eigenvalues
$\Omega_1,\Omega_2$ cross half-integer values. In
Table~\ref{tablek1k2} we report the values of $k_1$ and $k_2$ defined
from Eq. (\ref{analytick1k2}), according to the values of
$\Omega_1,\Omega_2$ computed at both Lagrangian points $L_1,L_2$ for
different values of $\mu$. The transitions in the values of $k_1,k_2$,
occurring when $\Omega_1,\Omega_2$ cross half integer values, are
highlighted in bold.

Hereafter we assume that the Floquet transformation ${\cal C}(f;e)$ is
defined with $k_1,k_2$ satisfying Eq.~(\ref{analytick1k2}).  In fact,
the definition of a close to the identity Floquet transformation is
essential in order to compute effectively the Birkhoff transformations
which remove from the Hamiltonian the dependence on $f$ up to an
higher order $N>2$: if the Floquet transformation is not close to the
identity, the Fourier expansion with respect to $f$ of the transformed
Hamiltonian, at any order $j$, contains a large number of terms of
large coefficients, and consequently the computation of the Birkhoff normal
forms saturates the computer memory at low normalization orders $N$.
\vskip 0.4 cm
\noindent
{\bf The Hamiltonian $\tilde H$ conjugate to $H$ by ${\cal C}(f;e)$.} Since 
the matrix ${\cal C}(f;e)$ is symplectic, the Floquet transformation 
$$
(\mathbf{q},\mathbf{p})={\cal C}(f;e)(\mathbf{\tilde q},\mathbf{\tilde p}) ,
$$
is canonical and conjugates the non-autonomous Hamiltonian 
$H(\mathbf{q},\mathbf{p},f;e)$ to the non-autonomous Hamiltonian:
\begin{equation}\label{tildeH}
\tilde H(\mathbf{\tilde q},\mathbf{\tilde p},f;e)=
\tilde H_2(\mathbf{\tilde q},\mathbf{\tilde p};e)+
\tilde H_3(\mathbf{\tilde q},\mathbf{\tilde p},f;e)+\ldots
\end{equation}
where from standard computations we have:
$$
\tilde H_2 = {1\over 2}\mathbf{y}\cdot (\mathbb{E}^T\hat B_e(k_1,k_2)) 
 \mathbf{y} \ \ \ \ ,\ \ \mathbf{y}=(\mathbf{\tilde q},\mathbf{\tilde p})
$$
is independent on $f$, and for $j\geq 3$ we have
$$
\tilde H_j= H_j({\cal C}(f;e)\mathbf{y}, f;e)  .
$$
Notice that, since the canonical transformation
is not autonomous, $\tilde H$ is not identified with 
$H({\cal C}(f;e)\mathbf{y}, f;e)$.


\vskip 0.4 cm
\noindent
For the explicit computation of the polynomials $\tilde
H_j$ it is convenient to represent the periodic matrix ${\cal C}(f;e)$
as a Fourier expansion:
\begin{equation}\label{FourierFloquet}
{\cal C}(f;e)={\mathbb I}+\sum_{\nu \in {\Bbb Z}}{\cal C}_\nu(e)e^{i\nu f}~.
\end{equation}
For practical purposes the series will be truncated and replaced
by a sum over all $\nu$ satisfying $\norm{\nu}\leq 2^{{\cal N}}$, 
for some convenient ${\cal N}$. 

The composition of each monomial of the Fourier-Taylor expansion of $H$:
\begin{displaymath}
a_{\nu,m_1,m_2,m_3} {\mathrm e}^{i\nu f}q_1^{m_1}q_2^{m_2}q_3^{m_3}\ \ ,\ \ 
\sum_{i=1}^3m_i=j\geq 3
\end{displaymath}
($a_{\nu,m_1,m_2,m_3}$ is a numerical coefficient depending only 
on $\mu,e$) with the Floquet transformation, 
provides a perturbation of
\begin{displaymath}
a_{\nu,m_1,m_2,m_3}  {\mathrm e}^{i\nu f}\tilde q_1^{m_1}\tilde q_2^{m_2}\tilde q_3^{m_3}
\end{displaymath}
which is represented as a Fourier--Taylor 
expansion of terms:
\begin{equation}\label{eq:newterms}
c_{\tilde \nu, \tilde m,\tilde n}(e)\, {\mathrm e}^{i \tilde \nu f}\, 
\tilde q_1^{\tilde m_1}\tilde q_2^{\tilde m_2}\tilde q_3^{\tilde m_3}\tilde p_1^{\tilde n_1}\tilde p_2^{\tilde n_2}\tilde p_3^{\tilde n_3}\ \ ,\ \ \sum_{i=1}^3(\tilde m_i+\tilde n_i)=j~,
\end{equation}
where the coefficient $c_{\tilde \nu, \tilde m,\tilde n}(e)$ is proportional to 
a product of $k\in [1,\ldots ,j]$ entries of the matrices 
${\cal  C}_{\nu^\star}(e)$ (with suitable $\nu^\star \in \mathbb{Z}$).   
\vskip 0.4 cm
Therefore, the Floquet Transformation increases significantly the Fourier-Taylor
expansion of the Hamiltonian. Since the convergence radius of Birkhoff
transformations depend not only on the resonance properties of the
linear frequencies, but also on the amplitudes of the coefficients of
the terms (\ref{eq:newterms}) of the Fourier-Taylor expansion of the
Hamiltonian, it is convenient to select a Floquet transformation
which is close to the identity, so that in all these terms 
the coefficients $c_{\tilde \nu, \tilde m,\tilde n}(e)$
are small for small values of the eccentricity. This property improves also 
the efficiency of the numerical computations of the Floquet-Birkhoff normal 
forms with a computer algebra system, since terms whose amplitude are smaller
than a threshold representing the numerical precision, are neglected.  

\vskip 0.4 cm
\noindent   
{\bf Proof of Lemma.} From elementary linear algebra there exists a real 
symplectic matrix $C$ 
conjugating the monodromy matrix $\mathbf{ \Phi_e}$ to 
the matrix
 \begin{equation}\label{eq:phitC}
 \mathbf{\tilde \Phi_e}:= C^{-1} \mathbf{\Phi_e }C =  \left(
  \begin{array}{c c c c c c}
 a_1    & 0  & 0 & \pm b_1  & 0 & 0 \\
    0 &  a_2 & 0  & 0 & \pm b_2  & 0 \\
    0 &  0 & e^{\pm 2\pi\lambda}  & 0 & 0 &    \\
    \mp b_1 & 0     & 0 &  a_1 & 0 & 0 \\
    0 &  \mp b_2 & 0  & 0 &  a_2 & 0 \\
    0 & 0  & 0 & 0 & 0 & e^{\mp 2\pi\lambda} \\    
  \end{array}
  \right) \ \ .
\end{equation}
The matrix $C$ as well as the signs in (\ref{eq:phitC}) are explicitly
determined by the eigenvectors of $\mathbf{\Phi_e }$. Let $\omega_1,\omega_2$ 
be defined as in (\ref{eq:omg}), and consider the matrix
$$
\tilde B(k_1,k_2)=  \left(
  \begin{array}{c c c c c c}
    0 &  0 & 0 & \pm(\omega_1+k_1) & 0 & 0 \\
    0 &  0 & 0 &   0 & \pm(\omega_2+k_2) & 0 \\
    0 &  0 & \pm\lambda & 0 & 0 & 0 \\
    \mp (\omega_1+k_1) & 0 & 0 & 0 & 0 & 0 \\
    0 &  \mp (\omega_2+k_2) & 0 & 0 & 0 & 0 \\
    0 & 0  & 0 & 0 & 0 & \mp \lambda \\    
  \end{array}
  \right)   
$$
where $k_1,k_2$ are arbitrary integer numbers, and the choice of the 
signs is done according to the signs appearing in (\ref{eq:phitC}). 
We have 
$$
 \mathrm{e}^{2\pi \tilde B(k_1,k_2)} = \mathbf{\tilde \Phi_e}~.
$$
Finally, since 
$\hat B(k_1,k_2)= C\tilde B(k_1,k_2)C^{-1}$ we obtain 
$$
\mathrm{e}^{2\pi\hat B(k_1,k_2)} = \mathrm{e}^{2\pi C\tilde B(k_1,k_2)C^{-1} }= 
C  \mathrm{e}^{2\pi \tilde B(k_1,k_2) }C^{-1} = \mathbf{\Phi_e }
$$ 
thus proving Eq. (\ref{eq:laB}). Equation (\ref{hamb}) as well 
as the symplecticity of the  Floquet transformation ${\cal C}(f;e)$ follow 
from elementary algebra. Finally, if $\hat B_0(k_1,k_2)=A_0$, we have 
$$
{\cal C}(f;0)= \mathrm{e}^{f A_0}\mathrm{e}^{-f \hat B_0(k_1,k_2)}={\mathbb I}~.
$$

\subsection{The Birkhoff transformations}

Let us consider the Hamiltonian: 
\begin{equation}
\tilde H (\mathbf{\tilde q},\mathbf{\tilde p},f;e) = \tilde H_2 (\mathbf{\tilde q},\mathbf{\tilde p};e)+ \tilde  H_3(\mathbf{\tilde q},\mathbf{\tilde p},f;e)+ \ldots 
\end{equation}
where each term $ \tilde  H_j$ is polynomial of degree $j$ 
in the variables $\mathbf{\tilde q},\mathbf{\tilde p}$ and periodic in $f$ with period $2\pi$, conjugate to Hamiltonian (\ref{eq:expandedH}) by the Floquet
transformation. We further apply the linear canonical transformation
\begin{equation}
 ( \mathbf{\tilde q}, \mathbf{\tilde p}) = {\cal D}(\mathbf{\hat Q}, \mathbf{\hat P})
\end{equation}
conjugating $\tilde H_2 (\mathbf{\tilde q},\mathbf{\tilde p};e)$ 
to the function $k_2$ defined in Eq.~(\ref{eq:k2}) 
(with $\sigma_1=\pm (\omega_1+k_1),\sigma_2=\pm (\omega_2+k_2)$
defined according to the eigenvalues and eigenvectors of the monodromy 
matrix $\mathbf{\Phi_e}$ as explained in Subsection 3.1) and we introduce  
the Birkhoff complex canonical variables 
\begin{equation}
( \mathbf{\hat Q}, \mathbf{\hat P}) = \hat {\cal D}(\mathbf{\hat q}, \mathbf{\hat p})\label{complexD}
\end{equation}
defined by
\begin{displaymath}
\hat Q_3=\hat q_3\ \ ,\ \ \hat P_3=\hat p_3\ \ ,\ \ 
\hat Q_j= \frac{{\hat q}_j+\mathrm{i}\,{\hat p}_j}{\sqrt{2}}~,\ \ 
\hat P_j= \frac{\mathrm{i}\,{\hat q}_j+{\hat p}_j}{\sqrt{2}}~,\,\,j=1,2  ,
\end{displaymath}
conjugating $k_2( \mathbf{\hat Q}, \mathbf{\hat P})$ to
\begin{equation}\label{eq:kk2}
\hat H_2( \mathbf{\hat q},\mathbf{\hat p})=
\mathrm{i}\, \sigma_1 \, {\hat q}_1 {\hat p}_1+
\mathrm{i} \, \sigma_2 \, {\hat q}_2 {\hat p}_2 +\lambda\, \hat q_3 \hat p_3   .
\end{equation}
The two linear transformations conjugate the Hamiltonian 
$\tilde H (\mathbf{\tilde q},\mathbf{\tilde p},f;e)$ to
\begin{equation}\label{eq:expham}
  \hat{H}(\hat{\mathbf{q}},\hat{\mathbf{p}},F,f) = F + 
\hat H_2(\hat{\mathbf{q}},\hat{\mathbf{p}}) + \sum_{j=1} \hat{H}_j(\hat{\mathbf{q}},\hat{\mathbf{p}},f;e)~,
\end{equation}
where the variable $F$, conjugate to $f$, has been introduced in order
to conveniently deal with an autonomous Hamiltonian and the terms
$\hat{H}_j$ for $j\geq 3$ are polynomials of degree $j$ in the
variables $\mathbf{\hat q},\mathbf{\hat p}$ and periodic in $f$
with period $2\pi$. The terms $\hat{H}_j$ with $j\geq 3$ are
represented as sum of monomials of the form
\begin{equation}\label{eq:genmono}
  a^{(j)}_{\nu,m_1,m_2,m_3,l_1,l_2,l_3} {\mathrm e}^{i\nu f}\,
  \hat{q}_1^{m_1} \hat{q}_2^{m_2} \hat{q}_3^{m_3} \hat{p}_1^{l_1}
  \hat{p}_2^{l_2} \hat{p}_3^{l_3}~,~~\sum_{i=1}^3 (m_i+l_i) = j~.
\end{equation}
Our objective now is twofold: on one hand, we aim to uncouple the
hyperbolic variables $\hat{q}_3,\hat{p}_3$ from the elliptic variables
$\hat{q}_1,\hat{p}_1$ and $\hat{q}_2,\hat{p}_2$ and, simultaneously,
to remove the explicit dependence of $\hat{H}$ on $f$ up to any
arbitrary finite order $N$. This is achieved if $\sigma_1,\sigma_2$
satisfy the non-resonance conditions:
\begin{displaymath}
j_1 \sigma_1+ j_2\sigma_2 +j_3 \ne 0 \ \ \ \ \forall (j_1,j_2,j_3)\in
{\Bbb Z}^3: \ \ \norm{j_1}+\norm{j_2}\in [1,N] ,\ \ j_3\in {\Bbb Z} ,
\end{displaymath}
with a close to the identity canonical transformation ${\cal C}_N$ conjugating 
the Hamiltonian
\eqref{eq:expham}, that now we identify as the initial Hamiltonian
$\hat{H}^{(2)}$, to a normal form Hamiltonian
\begin{equation}\label{eq:minormf}
\hat{H}^{(N)} = F + \sum_{j=2}^N K^{(N)}_j(\hat{\mathbf{q}},\hat{\mathbf{p}})+
\sum_{j\geq N+1} \hat{H}^{(N)}_j (\hat{\mathbf{q}},\hat{\mathbf{p}},f)
\end{equation}
where $K^{(N)}_j$ do not depend on $F,f$ and are 
polynomials of degree $j$ depending on $\hat{\mathbf{q}},\hat{\mathbf{p}}$
only through the products $\hat{q}_1\hat{p}_1, \hat{q}_2\hat{p}_2,
\hat{q}_3\hat{p}_3$, while $\hat{H}^{(N)}_j$ are polynomials of degree 
$j$ with coefficients depending periodically on $f$ with period $2\pi$. 

The canonical transformation ${\cal C}_N$ is constructed from
the composition of a sequence of 
$N-2$ elementary canonical Birkhoff transformations. Precisely,
we define the sequence of canonical transformations:
\begin{equation}\label{eq:totalC}
{\cal C}_J= {\cal C}_{\chi_{J}}\, \circ \, {\cal C}_{J-1}~, \ \ J=3,\ldots , N
\end{equation}
conjugating $\hat{H}:=\hat{H}^{(2)}$ to the intermediate
Floquet-Birkhoff normal form Hamiltonians:
\begin{equation}\label{eq:intermediateminormf}
\hat{H}^{(J)} := \hat{H}^{(J-1)}\circ {\cal C}_J =
 F + \sum_{j=2}^J K^{(J)}_j(\hat{\mathbf{q}},\hat{\mathbf{p}})+
\sum_{j\geq J+1} \hat{H}^{(J)}_j (\hat{\mathbf{q}},\hat{\mathbf{p}},f)
\end{equation}
with the property that $K^{(J)}_j$ do not depend on $F,f$ and are 
polynomials of degree $j$ depending on $\hat{\mathbf{q}},\hat{\mathbf{p}}$
only through the products $\hat{q}_1\hat{p}_1, \hat{q}_2\hat{p}_2,
\hat{q}_3\hat{p}_3$, while $\hat{H}^{(J)}_j$ are polynomials of degree 
$j$ with coefficients depending periodically on $f$ with period $2\pi$.

The transformation ${\cal C}_{2}$ is the identity while 
${\cal C}_{\chi_{J}}$ is the Hamiltonian flow at time $f=1$ of 
suitable generating functions $\chi_{J}$  defined from the coefficients of
$\hat{H}^{(J-1)}$. Below we describe the definition of the 
 generating functions $\chi_{J}$ and the steps required for the 
algorithmic computation of 
each canonical transformation ${\cal C}_N$
and Hamiltonian $\hat{H}^{(N)}$ using the Lie series method 
(for an introduction to the method, see \cite{LaPlata,Pisa})
and implemented with a computer algebra system in the examples
presented in this paper. We remark that, when using a computer algebra system, 
we need to set a cut off on the Fourier expansions with respect to the 
periodic variable $f$.  
\vskip 0.4 cm
For each $J\geq 3$  we assume that the Hamiltonian 
$\hat{H}^{(J-1)}$ and the canonical transformation ${\cal C}_{N-1}$ are
known, and we proceed as follows. 
\vskip 0.2 cm
\noindent
First, from $\hat{H}^{(J-1)}$ we compute the generating function ${\chi_{J}}$:
\begin{equation}\label{eq:genfunc-a}
  \chi_J = \hspace{-0.6cm}
    \sum_{\substack{ {m_j},{l_j}\in {\Bbb N}:\\ \sum_n
        (m_n+l_n)=N,\\ m_1 \ne l_1 \lor m_2 \ne l_2 \lor \\
        m_3 \ne l_3 \lor \nu \ne 0}} \hspace{-0.3cm}
    \frac{-a^{(J-1)}_{\nu,m_1,m_2,m_3,l_1,l_2,l_3}}{
      \mathrm{i}\,\sigma_1(l_1-m_1) + \mathrm{i}\,\sigma_2(l_2-m_2)
      + \lambda \, (l_3-m_3) + \mathrm{i}\, \nu} \,
    {\mathrm e}^{i\nu f}\,
    \hat{q}_1^{m_1} \hat{q}_2^{m_2} \hat{q}_3^{m_3} \hat{p}_1^{l_1}
    \hat{p}_2^{l_2} \hat{p}_3^{l_3} .
\end{equation}
Next, we compute the canonical transformation
\begin{displaymath}
 {\cal C}_{\chi_{J}}(\hat{\mathbf{q}}^{(J)},\hat{\mathbf{p}}^{(J)},F^{(J)},f^{(J)})=
(\hat{\mathbf{q}}^{(J-1)},\hat{\mathbf{p}}^{(J-1)},F^{(J-1)},f^{(J-1)})~,
\end{displaymath}
defined by the Hamiltonian flow of the generating function 
$\chi_J$ at time $f=1$. The transformation ${\cal C}_{\chi_{J}}$ is
explicitly represented as the Lie series
\begin{equation}\label{eq:theCchin}
 \zeta = e^{\,L_{\chi_J}} \zeta' := \zeta' + \{\zeta', \chi_J \}+
{1\over 2} \{ \{\zeta', \chi_J \}, \chi_J \}+\ldots  ~,
\end{equation}
where $L_{\chi_J} := \{\cdot, \chi_J \}$, and $\zeta, \zeta'$ denote
any couple of variables
$\hat{\mathbf{q}}^{(J-1)},\hat{\mathbf{q}}^{(J)}$,
$\hat{\mathbf{p}}^{(J-1)},\hat{\mathbf{p}}^{(J)}$ or
$F^{(J-1)},F^{(J)}$
  respectively.  The transformed Hamiltonian is
  computed as a Lie series as well:
\begin{equation}\label{eq:newham}
  \hat{H}^{(J)} = {\cal C}_{\chi_{J}}\, \hat{H}^{(J-1)}=
  e^{L_{\chi_J}}\, \hat{H}^{(J-1)} .
\end{equation}
The iteration ends for $J=N$, and finally, by reintroducing real
canonical variables,
\begin{equation}\label{eq:fromqtoQ}
  \begin{aligned}
  \hat{q}_1^{(N)} &= \frac{Q_1 - \mathrm{i}\, P_1}{\sqrt{2}}~, \quad
  &\hat{p}_1^{(N)} &= \frac{P_1 - \mathrm{i}\, Q_1}{\sqrt{2}}~,\\
  \hat{q}_2^{(N)} &=  \frac{Q_2 - \mathrm{i}\, P_2}{\sqrt{2}}~, \quad
  &\hat{p}_2^{(N)} &=\frac{P_2 - \mathrm{i}\, Q_2}{\sqrt{2}}~,\\
  \hat{q}_3^{(N)} &= Q_3~, \quad
 &\hat{p}_3^{(N)} &= P_3~,
  \end{aligned}
\end{equation}
and by suitably identifying the terms $k_j$ with $K_j$, and disregarding the dummy action $F^{(N)}$, we recover the final Floquet-Birkhoff normal form as
in Eq.~\eqref{eq:BTnf0} or~\eqref{eq:BTnf}.

\section{Experiments and examples}\label{sec:experiments}

\subsection{On the numerical computation of the Floquet-Birkhoff normal 
form}

In Section 3 we defined the Floquet-Birkhoff normal forms in the 
neighbourhood of a collinear equilibrium point of the ERTBP. For given 
values of the parameters $\mu,e$ the Floquet-Birkhoff normal
forms can be numerically provided as a Fourier-Taylor expansions:
\begin{displaymath}
a_{\nu,m_1,m_2,m_3}e^{\norm{\nu}} {\mathrm e}^{i\nu f}q_1^{m_1}q_2^{m_2}q_3^{m_3}\ \ ,\ \ \sum_{i=1}^3m_i=j\geq 3
\end{displaymath}
where the coefficients $a_{\nu,m_1,m_2,m_3}$ are floating point
numbers.  All the steps required to compute the Floquet-Birkhoff
normal form are explicit algebraic operations which can be implemented
with a computer algebra system. The only exception is the computation of
the principal fundamental matrix which demands the numerical
integration of a non-autonomous ODE.

In the present Subsection we provide an example of computation of the 
Floquet-Birkhoff normal form in a neighbourhood of the Lagrangian
point $L_1$ of the Earth-Moon ERTBP defined by $\mu = 0.0123$ and 
$e =0.0549006$; correspondingly we have $\beta=2.5764...$ (see Eq.
\eqref{eq:beta}).  The input Hamiltonian system is Hamiltonian 
\eqref{eq:expandedH0} explicitly computed as a Taylor expansion
in the variables $(\mathbf{q},\mathbf{p})$ up the polynomial order 
$N_{tot} = N + N_r $, where $N$ is the order of the Floquet-Birkhoff 
normal form, and $N_r\geq 1$ is needed to compute the lowest orders of 
the remainder. The examples below are computed for $N = 8$ and $N_r = 2$. 

\subsubsection{Computation of the canonical
Floquet transformation ${\cal C}(f;e)$}

{\bf Numerical computation of the principal  fundamental matrix solution.} 
The principal fundamental matrix solution $\Phi(f;e)$ of the linear
differential equation defined by the Hamilton equations of
$H_2$ (see Eq.~\eqref{eq:H2}):
\begin{equation}\label{eq:linsis}
  \left(
\begin{array}{c}
  \dot{\mathbf{q}}  \\
  \dot{\mathbf{p}}  
\end{array} \right)
= A(f;e) \, \left(
\begin{array}{c}
 \mathbf{q} \\
 \mathbf{p} 
\end{array} \right)
\end{equation}
can be provided as a Fourier series in the variable $f$ with floating
point coefficients, as it was done in~\cite{M05}.  The computation
requires the numerical integration of (\ref{eq:linsis}) with six
different initial conditions identified with the vectors
$\mathbf{u}^{(i)}\in {\Bbb R}^6$ of the standard basis of ${\Bbb
  R}^6$. Next, we apply the Fast Fourier Transform algorithm on the
outputs $\mathbf{u}^{(i)}(f)$ sampled on a regular grid of
values of
$$
f=f_j:=2\pi \, \frac{j }{2^{\cal N}}~,\ \ \ \ j=1,\ldots,2^{\cal N}~,
$$
where the value of ${\cal N} \in\mathbb{N}$ sets the Fourier cut-off in the 
variable $f$.

The implementation of these numerical procedures 
to the collinear Lagrangian points $L_1,L_2$ requires a careful 
check of the numerical precision, since the partially hyperbolic nature
of these equilibrium points determines an exponential loss of 
the numerical precision in the integration of the initial conditions
which may be responsible of large errors in the computation
of the matrices  $\Phi(f_j;e)$. For example, we find that the monodromy 
matrix $\mathbf{\Phi_e}$ 
has large entries of order $10^8$, and the characteristic polynomial
det$[\Phi_e -\lambda {\cal I}]$ has coefficients separated by 
8 orders of magnitude. To check the precision of the computation:
\begin{itemize}
\item[--] We perform the numerical integration of the linear 
equation \eqref{eq:linsis} with an explicit Runge-Kutta of order
six, quadruple floating point precision and integration
step $h =1 \snot[-4] \pi$. In order to check
the precision of our result we extend the computation over
the larger interval $[0,4\pi]$. 

\item[--] To prevent unnecessary
loss of precision digits due to the 
strong amplification of the norm of the solution vectors during the 
computation, we normalize the solution vector $\mathbf{x} = (\mathbf{q},\mathbf{p})$ every time its norm surpasses a certain threshold $\rho$, and we store in 
the computer memory the quantity $\Norm{\mathbf{x}}$; we continue the computation with the vector $\mathbf{x}/\Norm{\mathbf{x}}$. The threshold on the norm
of the solution vector that we used was $\rho = 1 \snot[3]$. Since the 
differential equation is linear, 
we reconstruct the solution $\mathbf{x}(f)$ by suitably multiplying 
the normalized solutions with the normalizing factors. This is the technique
introduced in~\cite{BGGS} for the precise numerical computation of 
the Lyapunov exponents. 

\item[--] Since the linear 
equation \eqref{eq:linsis} is periodic in $f$  of period $2\pi$, 
we check the precision of the numerical computation 
by checking if the eigenvalues of $\Phi(4\pi ;e)$ are the square of 
the eigenvalues of  $\Phi(2\pi ;e)$.  

\end{itemize}

Finally, the entries of the matrices $\Phi(f_j;e)$ are defined by
$$
\Phi(f_j;e)_{ki}= u^{(i)}_k(f_j)  .
$$
The experiments described below have been performed with ${\cal N} = 5$; we 
obtained (we here report only few precision digits): 
\begin{displaymath}
\mathbf{\Phi_e} = 
  \left(
  \begin{array}{c c c c c c}
    5.339 \snot[7] & 5.646 \snot[6] & 0 & 1.632\snot[7] & 7.725\snot[6] & 0 \\
    -2.556\snot[7] & -2.673 \snot[6] & 0 & -7.725\snot[6] & -3.657\snot[6] & 0 \\
    0 & 0 & -0.13223 & 0 & 0 & 0.44660 \\
    1.787 \snot[8] & 1.868 \snot[7] & 0 & 5.399 \snot[7] & 2.556\snot[7] & 0 \\
    -1.868\snot[7] & -1.953 \snot[6] & 0 & -5.646\snot[6] & -2.673\snot[6] & 0 \\
    0 & 0 & -2.19997 & 0 & 0 & -0.13223
  \end{array}
  \right)~,
\end{displaymath}
whose eigenvalues are $e^{(2\pi \lambda)} = 1.02644...\snot[8]$,
$e^{-(2\pi \lambda)} = 9.74245...\snot[-9]$, with $\lambda \sim 2.935896...$,
$a_1 \pm \mathrm{i}\, b_1 = -0.51780296... \pm \mathrm{i} \, 0.8554999...$ and $a_2 \pm \mathrm{i} b_2 = -0.132227... \pm \mathrm{i}\, 0.9912195...$.

\vskip 0.4 cm\noindent {\bf Choice of a logarithm of
  $\mathbf{\Phi_e}$.} The algebraic Lemma of Section 3 provides a
family of matrices $\hat B_e(k_1,k_2)$ such that, for any choice of
the integers $k_1,k_2$ the matrix $2\pi \hat B_e(k_1,k_2)$ is a
logarithm of the monodromy matrix $\mathbf{\Phi_e}$.  As already
remarked in Section 3, in order to obtain a close to the identity
Floquet transformation we chose $(k_1,k_2)=(2,2)$ according to
Eq. \eqref{analytick1k2}, with $\Omega_1 = 2.335547...$, $\Omega_2 =
2.270018...$ eigenvalues of the matrix $A_0$ (see
Eq.~\eqref{eq:laM}). From Eq.~\eqref{eq:laB} we have:
\begin{displaymath}
  \hat{B}_{e}(2,2) =
  \left(
  \begin{array}{c c c c c c}
    0 & 1.02669 & 0 & 1.03421 & 0 & 0 \\
    -1.03717 & 0 & 0 & 0 & 1.01949 & 0 \\
    0 & 0 & 0 & 0 & 0 & 1.02327 \\
    10.0729 & 0 & 0 & 0 & 1.03717 & 0 \\
    0 & -5.03017 & 0 & -1.02669 & 0 & 0 \\
    0 & 0 & -5.04063 & 0 & 0 & 0 
  \end{array}
  \right)~. 
\end{displaymath}

\begin{figure}[h]
      \centering
      \includegraphics[width=1.0\columnwidth]{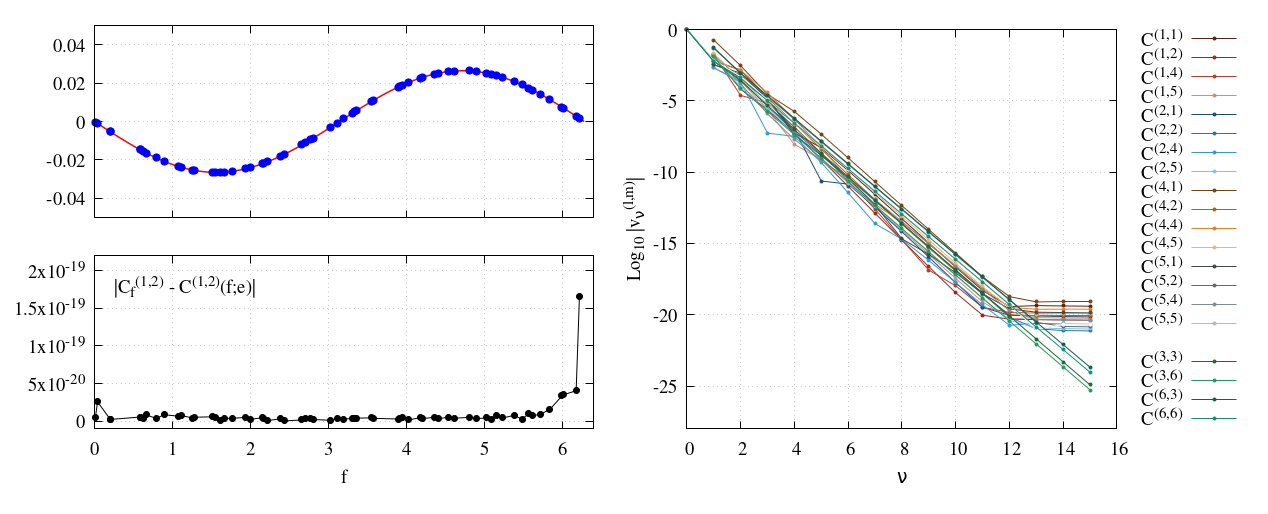}
      \caption{ Top left panel: comparison between the analytic expression
for ${\cal C}^{(1,2)}(f,e)$ provided by Eq.~\eqref{eq:flo2} (red line) 
and its Fourier expansion provided by Eq.~\eqref{eq:plm} computed on 
a random sample of values of $f$ (blue dots). In the bottom 
left panel we represent the absolute value of the difference between
the analytic expression for ${\cal C}^{(1,2)}(f,e)$ and the value of its 
Fourier expansion. In the right panel we provide 
all the values of $\norm{v_{\nu}^{(l,m)}}$ appearing in the Fourier
representations Eq.~\eqref{eq:plm} versus $\nu$.}
      \label{fig:accur}
    \end{figure}

\vspace{0.4cm}
\noindent
{\bf The Fourier series of ${\cal C}(f,e)$.} From Eq.~\eqref{eq:flo2} 
we obtain a sample of the matrix ${\cal C}(f,e)$:
\begin{equation}\label{eq:muchaspis}
C_{j} := {\cal C}(f_j,e)= \Phi(f_j,e) \, \mathrm{e}^{-\hat{B}_e(2,2,)\, f_j}~,\quad 
j=0,\ldots,2\pi
\end{equation}
which we use to compute a Fourier series for ${\cal C}(f,e)$, 
using the Fast Fourier Transform algorithm. Precisely, by denoting 
with $C_j^{(l,m)}$ the entries of $C_j$, the FFT algorithm gives
\begin{displaymath}
  \upsilon_{s}^{(l,m)} = \frac{1}{\sqrt{2^{{\cal N}}}} \sum_{r=1}^{2^{{\cal N}}} C_r^{(l,m)}
  \mathrm{e}^{2\pi\, \mathrm{i} (r-1)(s-1)/2^{{\cal N}}}~,
\end{displaymath}
which provide the Fourier representation:
\begin{equation}\label{eq:plm}
  {\cal C}^{(l,m)}(f,e) = \frac{1}{\sqrt{2^{{\cal N}}}} \sum_{s=0}^{2^{{\cal N}-1}-1}
  \upsilon_{s+1}^{(l,m)} \mathrm{e}^{- \mathrm{i}  s\,
    f} \frac{1}{\sqrt{2^{{\cal N}}}} + \sum_{s=2^{{\cal N}-1}}^{2^{{\cal N}}-1}
  \upsilon_{s+1}^{(l,m)} \mathrm{e}^{- \mathrm{i} 
    (s-2^{{\cal N}}) f}~.
\end{equation}
In Figure \ref{fig:accur} we plot a comparison between the values
of $C^{(1,2)}(f,e)$, computed using Eq. \eqref{eq:flo2}, and 
the values of the Fourier series compute using Eq. \eqref{eq:flo2}
on a random sample of values of $f$; we appreciate
that also for $f\ne f_j$ the difference sums to order $10^{-19}$. This 
number is in agreement with the decay of the values of the coefficients  
$\upsilon_{\nu}^{(l,m)}$, where the stabilization of the values of 
$\norm{\upsilon_{\nu}^{(l,m)}}$ at $10^{-19}$ for the largest $\nu$ provides
an indication that the numerical error is $10^{-19}$. The improvement of
this threshold value requires to increase the value of the Fourier cut-off
and to increase the numerical precision of the floating point arithmetics as well as of the numerical integration of Eq. \eqref{eq:linsis}.

\vspace{0.2cm}
\noindent
  
\noindent
{\bf Computation of the Fourier-Taylor expansion $\tilde{H} (\mathbf{\tilde
  q},\mathbf{\tilde p},f;e)$.} The canonical Floquet transformation conjugates
the Hamiltonian $H (\mathbf{q},\mathbf{p},f;e)$ to the Hamiltonian 
$\tilde{H} (\mathbf{\tilde  q},\mathbf{\tilde p},f;e)$. Since the 
Fourier decomposition of the ${\cal C}(f,e)$ is limited to 
the Fourier cut-off $2^{{\cal N}-1}$, we preliminary compute the 
Taylor-Fourier expansion of $H (\mathbf{q},\mathbf{p},f;e)$ 
by expanding the function
\begin{displaymath}
\frac{1}{1 + e\, \cos f} = \sum_{\nu \in {\Bbb Z}} \alpha_{\nu}
\mathrm{e}^{i \, \nu \, f}
\end{displaymath}
and by limiting the expansion to the Fourier cut-off $2^{{\cal N}-1}$. The
ultra-violet part of the Hamiltonian $H (\mathbf{q},\mathbf{p},f;e)$  
will be neglected. The second order term of the expansion is:
\begin{equation}\label{eq:tildeH2example}
  \begin{aligned}
  \tilde H_2 (\mathbf{\tilde q},\mathbf{\tilde p};e) =
 & \, 0.517196 \, \tilde{p_1}^2 
  + 0.509743 \, \tilde{p_2}^2 
  + 0.511635 \, \tilde{p_3}^2  
  - 1.03717 \, \tilde{p_2}\, \tilde{q_1} \\
&  + 1.02669   \, \tilde{p_1}\, \tilde{q_2} 
  - 5.03647  \tilde{q_1}^2
 + 2.51509 \, \tilde{q_2}^2
 + 2.52031 \, \tilde{q_3}^2
 \end{aligned}
\end{equation}
while the higher order terms $ \tilde H_j$ are polynomials with 
coefficients expanded as a Fourier series of $f$ with Fourier cut-off $2^{{\cal N}-1}$. 

\subsubsection{Computation of the second-order normal form}

To compute the linear canonical transformations \eqref{eq:linD2} 
and \eqref{complexD}
conjugating the second order Hamiltonian $\tilde H_2$ in \eqref{eq:tildeH2example} to its normal form (\ref{eq:kk2}) we proceed as usual by 
considering the matrix $A = \mathbb{E}\nabla \tilde{H}_2$, and
by computing its eigenvalues $\pm \mathrm{i} \sigma_1 =
\pm \mathrm{i}\, 2.336625$, $\pm \mathrm{i} \sigma_2 = \pm \mathrm{i}\,
2.271106$, $\pm \lambda = \pm 2.935895$, and the associated eigenvectors
${v_{-\sigma_1}},{v_{\sigma_1}},{v_{-\sigma_2}},{v_{\sigma_2}}\in {\Bbb C}^6$,
$v_{-\lambda},v_{\lambda}\in {\Bbb R}^6$. Then, we compute the 
coefficients  $c_1$, $c_2$, $c_3$ such that the matrix
$$
{\cal D}_0 = \left( c_1 {v_{\sigma_1}}, c_2
{v_{\sigma_2}}, c_3 {v_{\lambda}}, \mathrm{i}\,c_1
{v_{-\sigma_1}}, \mathrm{i}\, c_2 {v_{-\sigma_2}},
c_3 {v_{-\lambda}}\right)
$$
(i.e. the first column of ${\cal D}_0$ is the vector  $c_1 v_{\sigma_1}$, etc.)
is symplectic.
The linear transformation
\begin{equation}\label{eq:exampleD}
  \left(\begin{array}{c} 
    \tilde{\mathbf{q}}\\ 
    \tilde{\mathbf{p}} 
    \end{array} \right)
    = {\cal D}_0
  \left( \begin{array}{c}
    \hat{\mathbf{q}} \\
    \hat{\mathbf{p}} 
  \end{array} \right)
\end{equation}
is canonical and conjugates  the second order Hamiltonian $\tilde H_2$ in 
\eqref{eq:tildeH2example} to the second order normal form 
\begin{equation}\label{eq:hatH2example}
  \hat{H}_2 (\hat{\mathbf{q}},\hat{\mathbf{p}}) =
  2.336625 \, \mathrm{i} \, \hat{q}_1 \hat{p}_1
  + 2.271106 \, \mathrm{i} \, \hat{q}_2 \hat{p}_2
  + 2.935895 \, \hat{q}_3 \hat{p}_3~. 
\end{equation}
and for $j\geq 3$ the terms $ \tilde H_j$ to polynomials $\hat{H}_j$ of 
degree $j$, whose coefficients are periodic in $f$ and expanded as 
Fourier series with cut-off $2^{{\cal N}-1}$. The Hamiltonian:
\begin{equation}\label{eq:exphamexampl}
  \hat{H}(\hat{\mathbf{q}},\hat{\mathbf{p}},f,F;e) = F + 
\hat H_2(\hat{\mathbf{q}},\hat{\mathbf{p}}) + \sum_{j=1}^{N_{tot}} \hat{H}_j(\hat{\mathbf{q}},\hat{\mathbf{p}},f;e)~,
\end{equation}
where  the variable $F$, conjugate to $f$, has been introduced in 
order to conveniently deal with an autonomous Hamiltonian, is the 
input of the Birkhoff normalization algorithm defined in Subsection 3.2.

\subsubsection{The Floquet-Birkhoff normal form}

The Birkhoff  normalization can be implemented if for the specific values
of $\mu,e$ there are no
resonances: 
\begin{displaymath}
j_1 \sigma_1+ j_2\sigma_2 +j_3 \ne 0 \ \ \ \ \forall 
(j_1,j_2)\in {\Bbb Z}^3: \norm{j_1}+\norm{j_2}\in [1,N] , \forall
j_3\in {\Bbb Z}
\end{displaymath}
of order smaller or equal than $N=3$. For the values of $\mu,e$
indicated previously, we provide the details of the computation of the
Floquet-Birkhoff normal form of order $N=8$, by performing $N-2 = 6$
Birkhoff transformations defined in Subsection 3.2. Since the
computation of the Birkhoff transformations and of all the
intermediate Hamiltonians using the Lie series method are fully
described in Subsection 3.2, we here report the results. We find that
the Floquet-Birkhoff normal form Hamiltonian of order 8 is given by
\begin{equation}\label{eq:finalnormf}
\hat{H}^{(8)} = F + \sum_{j=1}^{J/2} K^{(8)}_{2j}(\hat{\mathbf{q}},\hat{\mathbf{p}})+
\sum_{j\geq 9} \hat{H}^{(8)}_j (\hat{\mathbf{q}},\hat{\mathbf{p}},f)
\end{equation}
where:
\begin{displaymath}
   \begin{aligned}
  K_2^{(8)}(\hat{\mathbf{q}},\hat{\mathbf{p}}) = & \hat{H}_2
  (\hat{\mathbf{q}},\hat{\mathbf{p}}) =2.336625\, \mathrm{i}\,
  \hat{q}_1 \, \hat{p}_1 + 2.271106\, \mathrm{i}\, \hat{q}_2 \,
  \hat{p}_2 + 2.935895\, \hat{q}_3 \, \hat{p}_3
  \\ K_4^{(8)}(\hat{\mathbf{q}},\hat{\mathbf{p}}) = & \, 7.076324 \,
  \hat{q}_1^2 \, \hat{p}_1^2 + 3.187254 \, \hat{q}_1 \, \hat{p}_1 \,
  \hat{q}_2 \, \hat{p}_2 + 6.326523 \, \hat{q}_2^2 \, \hat{p}_2^2
  \\ &\, - 32.88244 \, \mathrm{i}\, \, \hat{q}_1 \, \hat{p}_1\,
  \hat{q}_3 \, \hat{p}_3 - 30.07314 \, \mathrm{i}\, \hat{q}_2 \,
  \hat{p}_2 \, \hat{q}_3 \, \hat{p}_3 - 9.578629 \, \hat{q}_3^2 \,
  \hat{p}_3^2
  \end{aligned}
\end{displaymath}
and the coefficients of $K^{(8)}_6,K^{(8)}_8$ are reported in
Table~\ref{tab:nf} ($K^{(8)}_j$ do not depend on
$F,f$ and are polynomials of degree $j$ depending on
$\hat{\mathbf{q}},\hat{\mathbf{p}}$ only through the products
$\hat{q}_1\hat{p}_1, \hat{q}_2\hat{p}_2, \hat{q}_3\hat{p}_3$; the
notations of Table~\ref{tab:nf} is in agreement with
Eq.~\eqref{eq:genmono}).

The terms denoted with $\sum_{j\geq 9} \hat{H}^{(8)}_j$ are explicitly
computed for $j=9,10$, and are referred below as the remainder of the
Floquet-Birkhoff normal form. We find 97233 terms in in the remainder
with coefficients larger than $10^{-16}$.

\begin{figure}[h]
  \centering
  \includegraphics[width=1.\columnwidth]{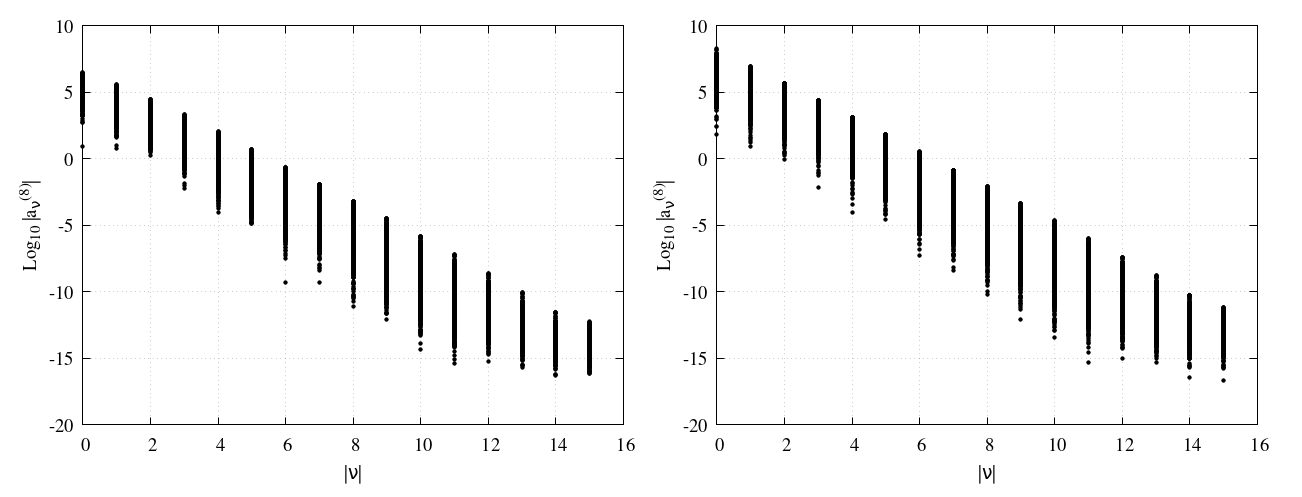}
  \caption{Logarithmic representation of the values of the
    coefficients $\norm{a^{(8)}_{\nu,m_1,m_2,m_3,l_1,l_2,l_3}}$ appearing
  in the terms $\hat{H}_{9}^{(8)}$ (left) and $\hat{H}_{10}^{(8)}$
  (right) of the remainder of the Hamiltonian~\eqref{eq:finalnormf}:
  for each value of $\nu$, we plot a dot corresponding to the
  logarithm of
  $\norm{a^{(8)}_{\nu,m_1,m_2,m_3,l_1,l_2,l_3}}$ for all
  the possible values of $m_j,l_j$.}
  \label{fig:remainder_decay}
\end{figure}

\begin{figure}[h]
  \centering
  \includegraphics[width=1.\columnwidth]{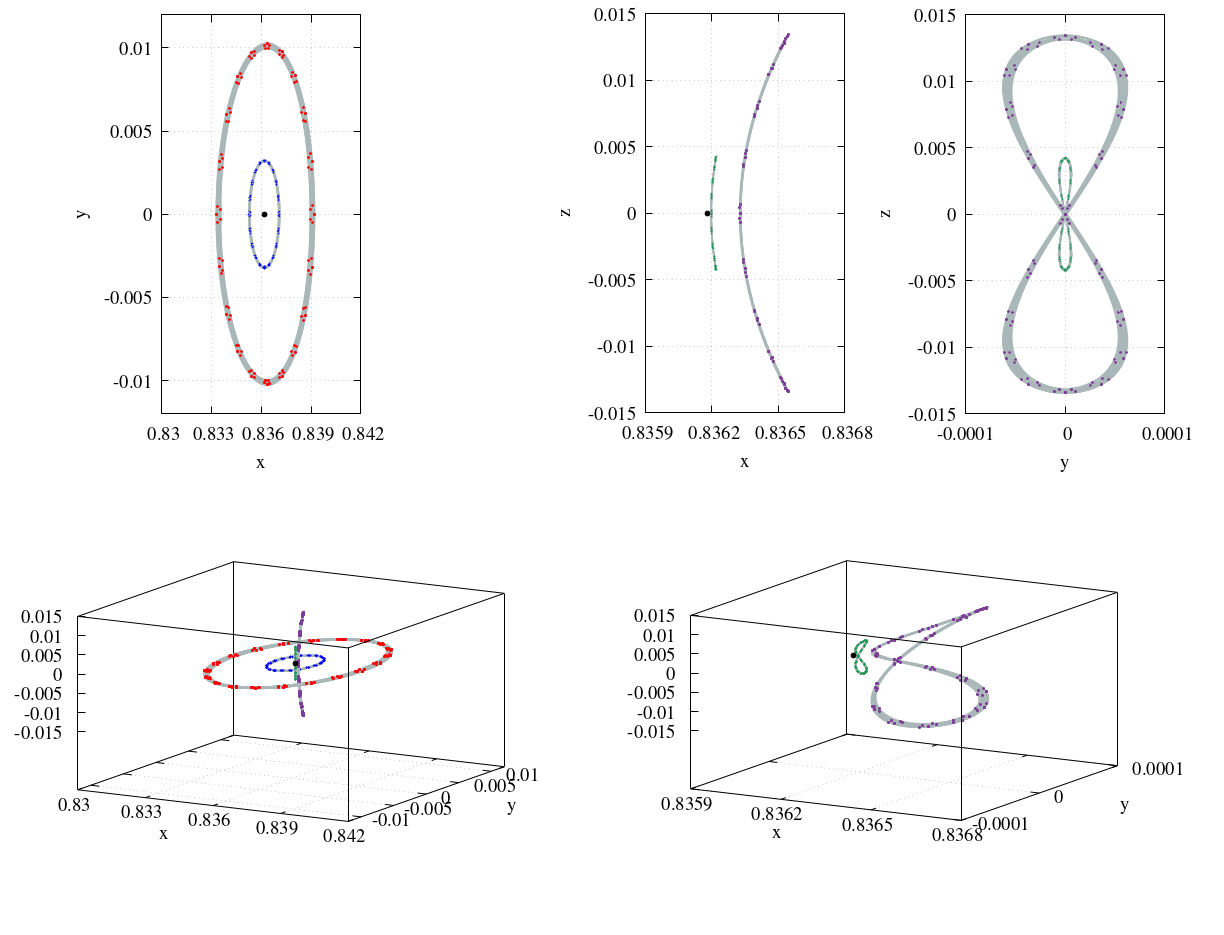}
  \caption{Representation of the points of the orbits used for the
    estimation of the norm of the remainder in
    Table~\ref{tab:rem}. The blue and red sets are in the planar
    two-dimensional tori ${\cal M}_{I_B,0}$ and ${\cal M}_{I_R,0}$
    respectively; the green and purple sets are in the vertical
    two-dimensional tori ${\cal M}_{0,I_G}$, ${\cal M}_{0,I_P}$. In all the
    cases, the complete orbits are represented in gray.}
  \label{fig:orbitrem}
\end{figure}

Figure \ref{fig:remainder_decay} provides a snapshot of the decay of
the Fourier harmonics $a^{(j)}_{\nu,m_1,m_2,m_3,l_1,l_2,l_3}$ with
$\nu$ (see Eq.~\eqref{eq:genmono}). The exponential decay of the
harmonics with the label $\nu$ appears clearly. As it is typical of
Birkhoff normal form remainders, the absolute values of the
coefficients increase with the order $j$, so that the convergence of
the remainder must be checked in neighbourhoods
$(\mathbf{\hat{q}},\mathbf{\hat{p}})=(0,0)$.  As a consequence, we
compute the maximum value of the norm of the remainder along bounded
orbits of the planar and of the spatial problems. The choice of the
initial conditions is done in the normalized variables
$(\hat{\mathbf{q}},\hat{\mathbf{p}})$ of the Floquet-Birkhoff normal
form \eqref{eq:exphamexampl} of order $N=8$. Precisely, we consider
the 4 sets of points $(\mathbf{\hat{q}},\mathbf{\hat{p}},f)$ in the
planar two-dimensional tori ${\cal M}_{I_B,0}$, ${\cal M}_{I_R,0}$ (the sets
{\it i},{\it ii}) and in the fully spatial two-dimensional tori
${\cal M}_{0,I_G}$, ${\cal M}_{0,I_P}$
(the sets {\it iii}, {\it iv}); the colors
refer to Figure \ref{fig:orbitrem}:

\vspace{0.4cm}
\noindent
{\it i.} Blue points: $\hat{q}_{1} = -\mathrm{i} \sqrt{I_B} \,
\mathrm{e}^{\mathrm{i} \phi}$, $\hat{p}_{1} = \sqrt{I_B } \,
\mathrm{e}^{\mathrm{i} \phi}$, with $I_B = 1\snot[-5]$ and $\phi = j
(2\pi/20)$, $f = i (2\pi/5)$, $j=1,20$, $i=1,5$; $\hat{q}_2 =
\hat{p}_2 = \hat{q}_{3} = \hat{p}_{3} = 0$, $\kappa_B = \hat{{\cal
    K}}(I_B,0,0) \approx 2.33655 \snot[-5]$.

\vspace{0.3cm}
\noindent
{\it ii.} Red points: $\hat{q}_{1} = -\mathrm{i} \sqrt{I_R} \,
\mathrm{e}^{\mathrm{i} \phi}$, $\hat{p}_{1} = \sqrt{I_R} \,
\mathrm{e}^{\mathrm{i} \phi}$, with $I_R = 1\snot[-4]$ and with $\phi
= j (2\pi/20)$, $f = i(2\pi/5)$, $j=1,20$, $i=1,5$; $\hat{q}_{2} =
\hat{p}_{2} = \hat{q}_{3} = \hat{p}_{3} = 0$, $\kappa_R = \hat{{\cal
    K}}(I_R,0,0) \approx 2.33655 \snot[-4]$.

\vspace{0.3cm}
\noindent
{\it iii.} Green points: $\hat{q}_{2} = -\mathrm{i} \sqrt{I_G} \,
\mathrm{e}^{\mathrm{i} \phi}$, $\hat{p}_{2} = \sqrt{I_G} \,
\mathrm{e}^{\mathrm{i} \phi}$, with $I_G = 2\snot[-5]$ and $\phi =
j(2\pi/20)$, $f = i(2\pi /5)$, $j=1,20$, $i=1,5$; $\hat{q}_{1} =
\hat{p}_{2} = \hat{q}_{1} = \hat{p}_{3} = 0$, $\kappa_G = \hat{{\cal
    K}}(0,I_G,0) \approx 4.54196 \snot[-5]$.

\vspace{0.3cm}
\noindent
{\it iv.} Purple points: $\hat{q}_{20} = -\mathrm{i} \sqrt{I_P} \,
\mathrm{e}^{\mathrm{i} \phi}$, $\hat{p}_{20} = \sqrt{I_P} \,
\mathrm{e}^{\mathrm{i} \phi}$, with $I_P = 2\snot[-4]$ and $\phi =
j(2\pi/20)$, $f = i(2\pi/5)$, $j=1,20$, $i=1,5$; $\hat{q}_{10} =
\hat{p}_{20} = \hat{q}_{10} = \hat{p}_{30} = 0$, $\kappa_P =
\hat{{\cal K}}(0,I_P,0) \approx 4.53968 \snot[-4]$.

\vspace{0.3cm} In Fig.~\ref{fig:orbitrem}, for the four sets of points
with the corresponding color, the full orbits are represented in
gray.

We compute the maximum of the norm of the remainder:
\begin{equation}\label{eq:normrem}
  \norm{R^{(J)}} :=
  \mathrm{Max}_{(\hat{\mathbf{q}},\hat{\mathbf{p}},f) \in {\cal S}}
  \sum_{j=J+1}^{10}
  \norm{\hat{H}_j^{(J)}(\hat{\mathbf{q}},\hat{\mathbf{p}},f)}
\end{equation}
where ${\cal S}$ represent the sets
of points (i), (ii), (iii) or (iv), for all the
normalization orders $J=2,\ldots,8$. The results are summarized in
Table~\ref{tab:rem}, and show the orders of magnitude of improvement
in the error of our best Floquet-Birkhoff normal form (of order
$J=8$) with respect to the classical Floquet approximation where no Birkhoff
transformation are implemented (corresponding to order $J=2$).

\begin{table}[h]
\centering
\begin{tabular}{|c|c|c|c|c|}
   \hline
   \hline
   $j$ & $m_1 = l_1$ & $m_2 = l_2$ & $m_3 = l_3$ & 
$a_{m_1,m_2,m_3}^{(8)}$ \\
   \hline
   6 & 3 & 0 & 0 & $25.117460\, \mathrm{i}$ \\
   6 & 2 & 1 & 0 & $-782.054619 \, \mathrm{i}$\\
   6 & 1 & 2 & 0 & $791.940455 \, \mathrm{i}$\\
   6 & 0 & 2 & 0 & $15.932649 \, \mathrm{i}$ \\
   6 & 2 & 0 & 1 & $47.958271 $\\
   6 & 1 & 1 & 1 & $223.182838 $\\
   6 & 0 & 2 & 1 & $14.202204 $\\
   6 & 1 & 0 & 2 & $-210.843893 \, \mathrm{i} $\\
   6 & 0 & 1 & 2 & $-141.046741 \, \mathrm{i} $\\
   6 & 0 & 0 & 3 & $-54.461156 $\\
   \hline
   8 & 4 & 0 & 0 & $-101.849178$ \\
   8 & 3 & 1 & 0 & $1.4081041 \snot[5]$ \\
   8 & 2 & 2 & 0 & $-3.6931581 \snot[5]$ \\
   8 & 1 & 3 & 0 & $1.0572474 \snot[5]$ \\
   8 & 0 & 4 & 0 & $-12.515592 $\\
   8 & 0 & 3 & 1 & $-289.061089 \, \mathrm{i}$ \\
   8 & 2 & 1 & 1 & $-6.8347427 \snot[3]\, \mathrm{i}$\\
   8 & 1 & 2 & 1 & $9.388619 \snot[3] \, \mathrm{i}$\\
   8 & 0 & 3 & 1 & $-663.967899 \, \mathrm{i}$\\
   8 & 0 & 2 & 2 & $-2.088688 \snot[3]$\\
   8 & 1 & 1 & 2 & $-4.705106 \snot[3]$\\
   8 & 0 & 2 & 2 & $-2.791412 \snot[3]$\\
   8 & 1 & 0 & 3 & $-2.607692 \snot[3] \, \mathrm{i}$\\
   8 & 0 & 1 & 3 & $-1.057350 \snot[3] \, \mathrm{i}$\\
   8 & 0 & 0 & 4 & $-558.96388$\\
   \hline
   \hline
\end{tabular}
\caption{Coefficients and combination of powers appearing in the terms
   of the normal form~\eqref{eq:finalnormf}, for a maximum polynomial
   expansion of order $N=8$.}
\label{tab:nf}
\end{table}

\begin{table}[h]
\centering
\begin{tabular}{|c|c|c|c|}
  \hline
  \hline
  \multicolumn{2}{|c|}{Blue set} & \multicolumn{2}{|c|}{Red set} \\
  \hline
  $J$ & $|R^{(J)}|$ & $J$ & $|R^{(j)}|$ \\
  \hline
  $2$ & $8.301112\snot[-8]$ & $2$ & $2.779487\snot[-6]$ \\
  $3$ & $1.710948\snot[-9]$ & $3$ & $1.761621\snot[-7]$ \\
  $4$ & $2.212756\snot[-11]$ & $4$ & $7.291894\snot[-9]$ \\
  $5$ & $4.045467\snot[-13]$ & $5$ & $4.224178\snot[-10]$ \\
  $6$ & $7.702234\snot[-15]$ & $6$ & $2.557286\snot[-11]$ \\
  $7$ & $1.646918\snot[-16]$ & $7$ & $1.737068\snot[-12]$ \\
  $8$ & $3.911953\snot[-18]$ & $8$ & $1.292694\snot[-13]$ \\
  \hline
  \hline
  \multicolumn{2}{|c|}{Green set} & \multicolumn{2}{|c|}{Purple set} \\
  \hline
  $J$ & $|R^{(J)}|$ & $J$ & $|R^{(J)}|$ \\
  \hline
  $2$ & $2.373759\snot[-7]$ &  $2$ & $8.139141\snot[-6]$ \\
  $3$ & $6.881388\snot[-9]$ &  $3$ & $7.177421\snot[-7]$ \\
  $4$ & $1.261467\snot[-10]$ & $4$ & $4.233325\snot[-8]$ \\
  $5$ & $3.262695\snot[-12]$ & $5$ & $3.473169\snot[-9]$ \\
  $6$ & $8.793931\snot[-14]$ & $6$ & $2.983974\snot[-10]$ \\
  $7$ & $2.661757\snot[-15]$ & $7$ & $2.872606\snot[-11]$ \\
  $8$ & $8.927986\snot[-17]$ & $8$ & $3.001279\snot[-12]$ \\
  \hline
  \hline
\end{tabular}
\caption{Results of the estimation of the norm of the
  remainder~\eqref{eq:normrem} for the four sets of points represented
  in Fig.~\ref{fig:orbitrem}.}
\label{tab:rem}
\end{table}

\subsection{Computation of transit orbits}\label{sec:transits}

An immediate application of the Floquet-Birkhoff normal form is the
computation of initial conditions of transit orbits. In this
Subsection, we discuss this application in the direction of
Figures~\ref{fig:kappa1}, \ref{fig:kappa2}, \ref{fig:tubos} and
\ref{fig:tubosconnec_new}, already presented in
Section~\ref{sec:overview}, where the initial conditions have been
chosen using the Floquet-Birkhoff normal form, and the orbits have
been obtained by numerically integrating the Hamiltonian
\eqref{eq:ori3bp}. These examples refer to the planar problem. As
already described in Section~\ref{sec:overview}, for all small values
of $\kappa$ we obtain the initial conditions of transit orbits
according to their position with respect to manifold tubes
$W^{s,loc}_\kappa,W^{u,loc}_\kappa$, determined by the sign of ${\cal
  I}_3=Q_3P_3>0$. The transits of Figures~\ref{fig:kappa1},
\ref{fig:kappa2}, \ref{fig:tubos} and \ref{fig:tubosconnec_new} have
been computed using the highest order normal form $\hat{H}^{(8)}$ that
we have computed, disregarding the remainder terms. The
transformations between the Cartesian variables
($\mathbf{q},\mathbf{p}$) to the final Floquet-Birkhoff variables are
computed explicitly up to order $8$ as well, and will be denoted below
by
$$
(\mathbf{Q},\mathbf{P})=\Psi^{-1}(\mathbf{q},\mathbf{p},f;e)\ \ \ \ ,\ \ \ \ 
(\mathbf{q},\mathbf{p})=\Psi(\mathbf{Q},\mathbf{P},f;e)
$$ 
respectively.  

For the planar problem it is more convenient to use the value ${\cal I}_1$
of the planar tori defined by ${\cal M}_{{\cal I}_1,0}$ 
as an independent choice, for which we compute the value $\kappa$ of
the local energy correspondingly. For that value of 
$\kappa$, we compute in Cartesian coordinates: the planar torus defined 
by ${\cal M}_{{\cal I}_1,0}$, i.e.
\begin{equation}
\bigcup_{f,\phi\in [0,2\pi]} \{ \Psi(\mathbf{Q},\mathbf{P},f;e):
(Q_1,P_1)=\sqrt{2 \,{\cal I}_1}(\sin \phi,\cos \phi) \ \ ,\ \ 
Q_2,P_2,Q_3,P_3=0\}  ,\label{torus2d}
\end{equation}
the transit orbits and the zero velocity curves. 

As a demonstration, in the panels (a) of Fig.~\ref{fig:kappa1} and
Fig.~\ref{fig:kappa2} we report in pink color a sample of the set
\eqref{torus2d} computed for $\kappa = 0.0000233655$, $0.0000467296$,
$0.0002335917$ (Fig.~\ref{fig:kappa1}) and $\kappa = 0.00232952$,
$0.0115030$, $0.0204374$ (Fig.~\ref{fig:kappa2}); the section of the
tori corresponding to $f=0$ is depicted in black in each case. The
initial conditions for the planar transit orbits have been chosen in
the normalized variables satisfying ${\cal I}_2=0$, ${\cal I}_3 = Q_3
\, P_3 > 0$ and ${\cal I}_1$ compatible with the fixed value $\kappa$
of the local energy (see the discussion in Section 2). Once the value
of ${\cal I}_3$ has been fixed, we choose $\norm{P_3(0)}\gg
\norm{Q_3(0)}$, so as to construct initial conditions in the close
vicinity of the stable manifold $W^{s,loc}_\kappa$. The transit orbits
in Fig.~\ref{fig:kappa1} and Fig.~\ref{fig:kappa2} have been obtained
for ${\cal I}_3= 1. \snot[-10]$ and
\begin{equation}\label{eq:inicondtrans}
  \begin{aligned}
    Q_1(0) &= 0~, \qquad &P_1(0) &= \sqrt{2 \,{\cal I}_1}~, \\
    Q_2(0) &= 0  ~, \qquad &P_2(0) &= 0~, \\
    Q_3(0) &= {\cal I}_{3}/P_3(0) ~, \qquad &P_3(0) &= 1 \snot[-4]~.
  \end{aligned}
\end{equation} 
Finally, using the direct transformation
$\Psi(\mathbf{Q},\mathbf{P},f;e)$ and fixing the value of $f$,
e.g. $f=0$, the initial condition \eqref{eq:inicondtrans} is mapped in
Cartesian variables. In Fig.~\ref{fig:kappa1} we show the orbits with
these initial conditions, obtained from the numerical integration of
the Hamilton's equations of~\eqref{eq:ori3bp}. In each case, we also
show the corresponding torus \eqref{torus2d} and the zero velocity
curves, which are obtained by solving numerically Eq. \eqref{zvs},
i.e. by computing numerically the level curves of the local energy
$\hat {\cal K}$, approximated at order $N=8$, for sample values of
$f$. The color scale in the orbits indicates the variation of the
local energy with respect to the initial value, exhibiting the
preservation of the local energy during the whole transition. We
notice that for the largest value of $\kappa$, as soon as the transit
orbit quits a neighbourhood of the torus \eqref{torus2d}, it reaches
distances from the Lagrangian point comparable to the distance of
$L_1$ to $P_2$. In \cite{PG20} we have shown that at these distances
the normal forms computed using the Cartesian variables looses
convergence, due to the gravitational singularity represented by
$P_2$. Therefore we do not represent the zero velocity curves in this
case, since only very close to the torus \eqref{torus2d} we expect a
good conservation of the local energy. As visual reference, we have
included in this panel the zero velocity curve obtained from the
Circular R3BP, for the corresponding value of $\mu$.

In Fig.~\ref{fig:tubos} we demonstrate
more extensively the correlation between the choice of the initial
conditions in the Floquet-Birkhoff normalized variables and the four
possible transit properties (two transit and two non-transit family of
orbits): the two families of initial conditions with ${\cal I}_3>0$
(red and green orbits) produce transit orbits, while the two family of
initial conditions with ${\cal I}_3<0$ produce orbits which
'bounce' back when they approach the planar torus. This behavior is
more clearly represented when we consider the projection of these
orbits in the original $xy$ variables. Again, the choice of the initial
conditions have been done in the Floquet-Birkhoff normalized
variables, and the numerical integrations have been done in the
Cartesian variable as explained above for Fig.~\ref{fig:kappa1} and
Fig.~\ref{fig:kappa2} (see caption of Fig.~\ref{fig:tubos} for the
initial conditions).


Figure~\ref{fig:tubosconnec_new} also show the effects caused of the
variation of the anomaly $f$ in the projection of the orbits to the
Cartesian space. We appreciate that the effect of the eccentricity,
through the variation of the anomaly in the terms of the
transformation, is to generate a small time-dependent
pulsation. Bottom left panel show with more detail such a pulsation on
the ${\cal M}_{{\cal I}_1,0}$. A similar effect
takes place when we consider the projection of the transit orbits. In
these plots, we include also a family of orbits in the stable and
unstable manifolds of ${\cal M}_{{\cal I}_1,0}$ (gray orbits).
The initial conditions for the orbits in these manifold tubes
have been chosen in the Floquet-Birkhoff normalized variables, by
setting
\begin{equation}\label{eq:stabletubeic}
\bigcup_{f,\phi\in [0,2\pi]} \{ \Psi(\mathbf{Q},\mathbf{P},f;e):
(Q_1,P_1)=\sqrt{2 \,{\cal I}_1}(\sin \phi,\cos \phi)~,
\,\, P_3 \neq 0~, \,\,
Q_2,P_2,Q_3=0\}
\end{equation}
for the stable manifold tube, and
\begin{equation}\label{eq:unstabletubeic}
\bigcup_{f,\phi\in [0,2\pi]} \{ \Psi(\mathbf{Q},\mathbf{P},f;e):
(Q_1,P_1)=\sqrt{2 \,{\cal I}_1}(\sin \phi,\cos \phi)~,
\,\, Q_3 \neq 0~, \,\,
Q_2,P_2,P_3=0\}
\end{equation}
for the unstable manifold tube.

\begin{figure}[h]
  \centering
  \includegraphics[width=1.\columnwidth]{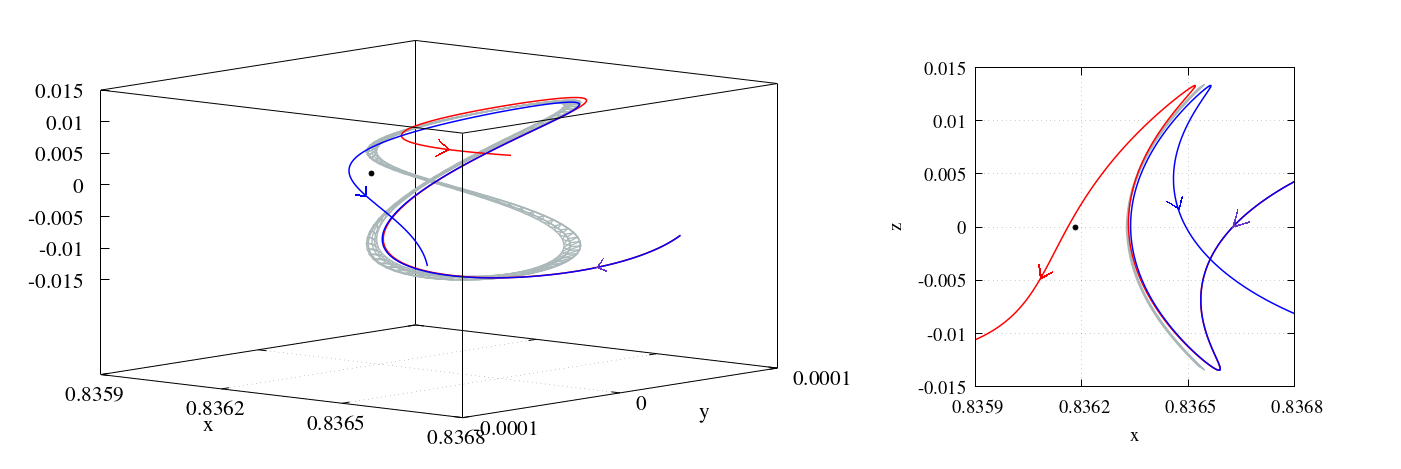}
  \caption{Two orbits of the spatial problem with very close initial
    conditions, but different transit properties. In the left-panel we
    represent the orbits in the Cartesian 3-dimensional $xyz$ space,
    while in the right-panel we represent the projection of the orbits
    in the Cartesian 2-dimensional $xz$ plane.
    The initial conditions are:
    $Q_1,P_1=0$, $Q_2=0$, $P_2= 1/50$, and $f = 0$ for the two
    orbits, while $Q_3 = -1\snot[-6]$, $P_3 = 1\snot[-4]$ for the
    blue orbit and $Q_3 =1\snot[-6]$, $P_3 = 1\snot[-4]$ for the red orbit.
    The gray orbit corresponds at the section $f=0$ of the 
    torus ${\cal M}_{0,{\cal I}_2,}$.The arrows indicate the
    direction of the motion for increasing values of $f$.}
  \label{fig:vertt}
\end{figure}

In Fig. \ref{fig:vertt} we finally provide an example of transit
orbits in the genuine spatial problem. The choice of the initial
conditions is done exactly as for the planar transit
orbits~\eqref{eq:inicondtrans}, except that we set ${\cal I}_1=0$ and
${\cal I}_2>0$, ${\cal I}_3>0$. The value of $\kappa$ is the largest
value considered for the orbits of Fig.~\ref{fig:orbitrem},
i.e. $\kappa(I_P) = 4.53968 \snot[-4]$.

\section{Conclusions}

The transits through the Lagrangian points of the circular restricted
three-body problem are relevant for the dynamics of comets and
spacecrafts.  To use the results obtained for the CRTBP in a realistic
model of the Solar System requires to take into account the elliptic
orbit of the planet of the close encounter, as well as the
perturbations from the other planets. Despite the eccentricity of the
planets is small, the ERTBP represents a major modification of the
CRTBP, since non global first integral are known, and the definition
of realms of admissible or forbidden motion and of the zero velocity
curves is lost.  Nevertheless the Lagrange solutions exist for both
problems, and using a combination of the Floquet theory and of
Birkhoff normalizations we have been able to recover a classification
of the transits occurring at the Lagrangian points $L_1,L_2$. We have
shown that an improvement of the traditional Floquet theory is indeed
possible, except for few values of the reduced mass $\mu$
corresponding to resonances.  These methods allow a full control of
the effect to the true anomaly $f$ (to use as a parameter) in the
classification of the transits, and provide an analytic way to
construct, for example, patched orbits more reliable than those of the
bi-circular models. This is left for future works, as well as an
analysis of the Arnold diffusion due to the remainder of Birkhoff
normal forms, along the lines of paper~\cite{GEP}.

\section*{Acknowledgments}

The authors acknowledge the project MIUR-PRIN 20178CJA2B "New frontiers
of Celestial Mechanics: theory and applications".



\end{document}